\documentclass[lettersize,journal]{IEEEtran}
\usepackage{amsmath,amsfonts}
\usepackage{amsthm}
\usepackage{algpseudocode}
\usepackage{algorithm}
\usepackage{array}
\usepackage[caption=false,font=normalsize,labelfont=sf,textfont=sf]{subfig}
\usepackage{textcomp}
\usepackage{stfloats}
\usepackage{url}
\usepackage{verbatim}
\usepackage{graphicx}
\usepackage{cite}
\usepackage{multirow}
\usepackage{hyperref}
\usepackage{mathrsfs}
\usepackage[printonlyused]{acronym}

\usepackage{color}

\usepackage[table]{xcolor}
\usepackage{makecell}
\usepackage{hhline}
\usepackage{booktabs}

\newcommand{\hthickline}{\noalign{\hrule height 0.9pt}}
\usepackage{tikz} 
\usepackage[utf8]{inputenc}
\usepackage{pgfplots} 
\makeatletter
\newcommand{\gettikzxy}[3]{%
\tikz@scan@one@point\pgfutil@firstofone#1\relax
  \edef#2{\the\pgf@x}%
  \edef#3{\the\pgf@y}%
}
\usepackage{units} % use \unit command
\usepackage{bm} % for bold symbol
\usepackage{amssymb} % for AMS symbols

\newtheorem{remark}{Remark}
 
\begin{document}
% transpose and hermitian
\newcommand{\TT}{\mathsf{T}}
\newcommand{\HH}{\mathsf{H}}

% Vectors
\newcommand{\av}{{\bf a}}
\newcommand{\bv}{{\bf b}}
\newcommand{\cv}{{\bf c}}
\newcommand{\dv}{{\bf d}}
\newcommand{\ev}{{\bf e}}
\newcommand{\fv}{{\bf f}}
\newcommand{\gv}{{\bf g}}
\newcommand{\hv}{{\bf h}}
\newcommand{\iv}{{\bf i}}
\newcommand{\jv}{{\bf j}}
\newcommand{\kv}{{\bf k}}
\newcommand{\lv}{{\bf l}}
\newcommand{\mv}{{\bf m}}
\newcommand{\nv}{{\bf n}}
\newcommand{\ov}{{\bf o}}
\newcommand{\pv}{{\bf p}}
\newcommand{\qv}{{\bf q}}
\newcommand{\rv}{{\bf r}}
\newcommand{\sv}{{\bf s}}
\newcommand{\tv}{{\bf t}}
\newcommand{\uv}{{\bf u}}
\newcommand{\wv}{{\bf w}}
\newcommand{\vv}{{\bf v}}
\newcommand{\xv}{{\bf x}}
\newcommand{\yv}{{\bf y}}
\newcommand{\zv}{{\bf z}}
\newcommand{\zerov}{{\bf 0}}
\newcommand{\onev}{{\bf 1}}
\newcommand{\avr}{\av_\text{R}}
% Matrices
\newcommand{\Am}{{\bf A}}
\newcommand{\Bm}{{\bf B}}
\newcommand{\Cm}{{\bf C}}
\newcommand{\Dm}{{\bf D}}
\newcommand{\Em}{{\bf E}}
\newcommand{\Fm}{{\bf F}}
\newcommand{\Gm}{{\bf G}}
\newcommand{\Hm}{{\bf H}}
\newcommand{\Id}{{\bf I}}
\newcommand{\Jm}{{\bf J}}
\newcommand{\Km}{{\bf K}}
\newcommand{\Lm}{{\bf L}}
\newcommand{\Mm}{{\bf M}}
\newcommand{\Nm}{{\bf N}}
\newcommand{\Om}{{\bf O}}
\newcommand{\Pm}{{\bf P}}
\newcommand{\Qm}{{\bf Q}}
\newcommand{\Rm}{{\bf R}}
\newcommand{\Sm}{{\bf S}}
\newcommand{\Tm}{{\bf T}}
\newcommand{\Um}{{\bf U}}
\newcommand{\Wm}{{\bf W}}
\newcommand{\Vm}{{\bf V}}
\newcommand{\Xm}{{\bf X}}
\newcommand{\Ym}{{\bf Y}}
\newcommand{\Zm}{{\bf Z}}
\newcommand{\Onem}{{\bf 1}}
\newcommand{\Zerom}{{\bf 0}}
% text uppercase
\newcommand{\At}{{\rm A}}
\newcommand{\Bt}{{\rm B}}
\newcommand{\Ct}{{\rm C}}
\newcommand{\Dt}{{\rm D}}
\newcommand{\Et}{{\rm E}}
\newcommand{\Ft}{{\rm F}}
\newcommand{\Gt}{{\rm G}}
\newcommand{\Ht}{{\rm H}}
\newcommand{\It}{{\rm I}}
\newcommand{\Jt}{{\rm J}}
\newcommand{\Kt}{{\rm K}}
\newcommand{\Lt}{{\rm L}}
\newcommand{\Mt}{{\rm M}}
\newcommand{\Nt}{{\rm N}}
\newcommand{\Ot}{{\rm O}}
\newcommand{\Pt}{{\rm P}}
\newcommand{\Qt}{{\rm Q}}
\newcommand{\Rt}{{\rm R}}
\newcommand{\St}{{\rm S}}
\newcommand{\Tt}{{\rm T}}
\newcommand{\Ut}{{\rm U}}
\newcommand{\Wt}{{\rm W}}
\newcommand{\Vt}{{\rm V}}
\newcommand{\Xt}{{\rm X}}
\newcommand{\Yt}{{\rm Y}}
\newcommand{\Zt}{{\rm Z}}

% Bold greek letters
\newcommand{\alphav}{\hbox{\boldmath$\alpha$}}
\newcommand{\betav}{\hbox{\boldmath$\beta$}}
\newcommand{\gammav}{\hbox{\boldmath$\gamma$}}
\newcommand{\deltav}{\hbox{\boldmath$\delta$}}
\newcommand{\etav}{\hbox{\boldmath$\eta$}}
\newcommand{\lambdav}{\hbox{\boldmath$\lambda$}}
\newcommand{\kappav}{\hbox{\boldmath$\kappa$}}
\newcommand{\epsilonv}{\hbox{\boldmath$\epsilon$}}
\newcommand{\nuv}{\hbox{\boldmath$\nu$}}
\newcommand{\muv}{\hbox{\boldmath$\mu$}}
\newcommand{\zetav}{\hbox{\boldmath$\zeta$}}
\newcommand{\phiv}{\hbox{\boldmath$\phi$}}
\newcommand{\psiv}{\hbox{\boldmath$\psi$}}
\newcommand{\thetav}{\hbox{$\boldsymbol\theta$}}
\newcommand{\tauv}{\hbox{\boldmath$\tau$}}
\newcommand{\omegav}{\hbox{\boldmath$\omega$}}
\newcommand{\xiv}{\hbox{\boldmath$\xi$}}
\newcommand{\sigmav}{\hbox{\boldmath$\sigma$}}
\newcommand{\piv}{\hbox{\boldmath$\pi$}}
\newcommand{\rhov}{\hbox{\boldmath$\rho$}}

\newcommand{\Gammam}{\hbox{\boldmath$\Gamma$}}
\newcommand{\Lambdam}{\hbox{\boldmath$\Lambda$}}
\newcommand{\Deltam}{\hbox{\boldmath$\Delta$}}
\newcommand{\Sigmam}{\hbox{\boldmath$\Sigma$}}
\newcommand{\Phim}{\hbox{\boldmath$\Phi$}}
\newcommand{\Pim}{\hbox{\boldmath$\Pi$}}
\newcommand{\Psim}{\hbox{\boldmath$\Psi$}}
\newcommand{\psim}{\hbox{\boldmath$\psi$}}
\newcommand{\chim}{\hbox{\boldmath$\chi$}}
\newcommand{\omegam}{\hbox{\boldmath$\omega$}}
\newcommand{\Thetam}{\hbox{\boldmath$\Theta$}}
\newcommand{\Omegam}{\hbox{\boldmath$\Omega$}}
\newcommand{\Xim}{\hbox{\boldmath$\Xi$}}
 % include macro 
\newacro{ft} [FT] {Fourier transform}
\newacro{mimo} [MIMO] {multiple-input multiple-output}
\newacro{ue} [UE] {user equipment}
\newacro{leo} [LEO] {low-Earth orbit}
\newacro{meo} [MEO] {medium Earth orbit}
\newacro{geo} [GEO] {geostationary Earth orbit}
\newacro{aod} [AoD] {angle-of-departure}
\newacro{aoa} [AoA] {angle-of-arrival}
\newacro{awgn} [AWGN] {additive white Gaussian noise}
\newacro{3d} [3D] {three-dimensional}
\newacro{los} [LOS] {line-of-sight}
\newacro{ofdm} [OFDM] {orthogonal frequency-division multiplexing}
\newacro{raan} [RAAN] {right ascension of the ascending node}
\newacro{gnss} [GNSS] {global navigation satellite systems}
\newacro{eci} [ECI] {Earth-centered inertial}
\newacro{ecef} [ECEF] {Earth-centered, Earth-fixed}
\newacro{tle} [TLE] {two-line-element}
\newacro{sgp4} [SGP4] {simplified general perturbation version 4}
\newacro{nlos} [NLOS] {non-line-of-sight}
\newacro{mcrb} [MCRB] {misspecified Cramér-Rao bound}
\newacro{crb} [CRB] {Cramér-Rao bound}
\newacro{mle} [MLE] {maximum likelihood estimation}
\newacro{ntn} [NTN] {non-terrestrial network}
\newacro{psd} [PSD] {power spectral density}
\newacro{fim} [FIM] {Fisher information matrix}
\newacro{6g} [6G] {6th Generation}
\newacro{5g} [5G] {5th Generation}
\newacro{mse} [MSE] {mean square error}
\newacro{lbm} [LBM] {lower bound matrix}
\newacro{lb} [LB] {lower bound}
\newacro{ris} [RIS] {reconfigurable intelligent surface}
\newacro{ls} [LS] {least squares}
\newacro{cp} [CP] {cyclic prefix}
\newacro{rms} [RMS] {root-mean-square}
\newacro{rmse} [RMSE] {root-mean-square error}
\newacro{3gpp} [3GPP] {3rd Generation Partnership Project}
\newacro{ipac} [IPAC] {integrated positioning and communication}
\newacro{pnt} [PNT] {positioning, navigation, and timing}
\newacro{rf} [RF] {radio-frequency}
\newacro{soop} [SoOP] {signals-of-opportunity}
\newacro{sqp} [SQP] {sequential quadratic programming}
\newacro{qp} [QP] {quadratic programming}
\newacro{db} [dB] {decibel}
\newacro{bs} [BS] {base station}
\newacro{enu} [ENU] {East–North–Up}
\newacro{snr} [SNR] {signal-to-noise ratio}
\newacro{norad} [NORAD] {North American Aerospace Defense Command}
\newacro{map} [MAP] {maximum a posteriori}
\newacro{sl} [SL] {satellite local}
\newacro{ml} [ML] {maximum likelihood}
 % include acronyms
% ------------------------------------------------------------------------------
\title{Orbital Errors in LEO Satellite Positioning: Modeling, Analysis, and Bayesian Calibration}

\author{Jie~Ma, Pinjun~Zheng, Xing~Liu, Yuchen~Zhang, Ali~A.~Nasir, and~Tareq~Y.~Al-Naffouri,~\IEEEmembership{Fellow,~IEEE}

\thanks{This publication is based upon work supported by King Abdullah University of Science and Technology (KAUST) under Award No. ORFS-CRG12-2024-6478.

Jie~Ma, Yuchen~Zhang, and~Tareq~Y.~Al-Naffouri are with the Electrical and Computer Engineering Program, Computer, Electrical and Mathematical Sciences and Engineering (CEMSE), King Abdullah University of Science and Technology (KAUST), Thuwal 23955-6900, Kingdom of Saudi Arabia (e-mails: \{jie.ma; yuchen.zhang; tareq.alnaffouri\}@kaust.edu.sa).

Pinjun~Zheng is with the School of Engineering, The University of British Columbia, Kelowna, BC V1V 1V7, Canada (e-mail: pinjun.zheng@ubc.ca). The majority of his contributions to this work were made during his Ph.D. studies at KAUST, Thuwal 23955-6900, Kingdom of Saudi Arabia.

Xing~Liu is with the Institut d'Estudis Espacials de Catalunya (IEEC), Barcelona, Spain (e-mail: xing@ieec.cat).

Ali~A.~Nasir is with the Department of Electrical Engineering and Center for Communication Systems and Sensing, King Fahd University of Petroleum and Minerals, Dhahran 31261, Kingdom of Saudi Arabia (e-mail: anasir@kfupm.edu.sa).
}

}

\maketitle

\begin{abstract}
Low-Earth orbit (LEO) satellites offer a promising alternative to global navigation satellite systems for precise positioning. However, their relatively low altitudes make them more susceptible to orbital perturbations, which in turn degrade positioning accuracy. In this work, we study (i) the mechanisms through which orbital errors affect positioning performance and~(ii) the fundamental calibration strategies for mitigating LEO orbital errors. We first model the satellite orbit and analyze historical two-line element data to characterize the statistical behavior of LEO orbital errors. Based on these real-data statistics, we derive the misspecified Cram\'er-Rao bound to quantify the impact of orbital errors on positioning performance. Subsequently, we develop a Bayesian orbit calibration framework that leverages the learned orbital error statistics as prior information to enhance calibration accuracy. This work sheds light on orbital error modeling, real-data-based evaluation, theoretical impact analysis, and calibration methodologies. Extensive simulations show that stale orbital information can severely degrade positioning performance, whereas learning and incorporating orbital error statistics can effectively enhance orbit calibration and ultimately improve positioning accuracy.

\end{abstract}


\section{Introduction}
	As highlighted in the \ac{3gpp} TR 38.811~\cite{3GPP38811} and TR 22.870~\cite{3GPP22870}, \acp{ntn} are identified as key components of ubiquitous connectivity for the \ac{5g} and \ac{6g} wireless networks. Within \acp{ntn}, \ac{leo} constellations have emerged as a promising platform for providing global positioning services. Unlike \ac{gnss}, which are deployed in \ac{meo}, \ac{leo} satellites operate at much lower altitudes, ranging from approximately \unit[300]{km} to \unit[2,000]{km}. The lower altitude leads to a shorter propagation distance and faster satellite motion, both of which bring potential advantages for positioning. The reduced propagation distance leads to stronger received signals and thus improves the quality of observable extraction~\cite{ferre2022leo,ferre2022feasibility}. Meanwhile, the faster satellite motion produces more pronounced Doppler dynamics, which provide useful information for positioning and velocity-related estimation. Another key advantage lies in the large-scale deployment of LEO constellations. Today, the dense deployment of \ac{leo} constellations, such as Starlink, OneWeb, and Project Kuiper, offers higher satellite visibility, richer geometric diversity, and greater measurement redundancy~\cite{ettefagh2025integrated,bader2025enhanced}, providing a strong foundation for next-generation positioning systems.

    Despite these advantages, a common challenge across \ac{leo} positioning systems is the inaccuracy of the available orbital information. Compared with \ac{meo} and \ac{geo} constellations, \ac{leo} satellites suffer from more pronounced orbital perturbations caused by atmospheric drag, the Earth's non-uniform gravitational field, and solar radiation pressure~\cite{stock2024survey}. The orbits of \ac{leo} satellites are tracked and can be accessed in the form of publicly released \ac{tle} sets, where each space object is identified by a \ac{norad} catalog number~\cite{KelsoGPFormats}. However, these \ac{tle} data are not updated in real time. Therefore, satellite states at arbitrary epochs are typically predicted using an orbit propagator such as the \ac{sgp4} model~\cite{Vallado2006SGP4}, potentially leading to non-negligible orbital errors. 
    
        \subsection{Related Works}
    Numerous studies have investigated the impact of orbital errors on \ac{leo}-based positioning and developed calibration methods, as reviewed below. 
    
    \subsubsection{Impact Evaluation of Orbital Errors}
    Orbital errors have been recognized as one of the main factors limiting the accuracy of \ac{leo}-based positioning. Existing studies have examined the impact of orbital errors on \ac{leo}-based positioning through both numerical simulations and practical experiments. Large-scale constellation simulations have shown that typical \emph{satellite orbital user-range errors} can induce meter-level positioning accuracy degradation~\cite{isprs-archives-XLVIII-1-W2-2023-1111-2023}. Complementary real-world experiments have reported even larger positioning errors, exceeding one kilometer along a single axis~\cite{du2024leo}. Related analyses have further indicated that differential schemes are also sensitive to \ac{leo} orbit errors~\cite{zhao2023analysis}. These studies demonstrate that inaccurate orbital information can lead to non-negligible positioning degradation.

    Although these studies reveal the practical importance of orbital errors, most of them focus on quantifying the final positioning degradation under specific simulation settings, experimental conditions, or observable models. The underlying impact mechanism, i.e., how orbital mismatch propagates through different geometric observables and eventually affects the positioning estimate, has not been fully analyzed. Moreover, existing evaluations usually treat orbital errors as given perturbations or equivalent measurement errors, with little attention paid to their statistical characterization and analysis. 
   
    \subsubsection{Orbit Calibration and Compensation Methodologies}
    Beyond impact evaluation, various calibration and compensation methods have been proposed to mitigate the effect of orbital errors on \ac{leo}-based positioning. These works can be roughly categorized into two types: (i) model-based refinement and (ii) learning-based calibration. 
    
    Model-based refinement mainly extends fundamental positioning frameworks by incorporating refined observation models or additional information with orbital error awareness. For instance, a Doppler-domain formulation in~\cite{wang2023doppler} introduced an \emph{orbit error equivalent Doppler measurement error (OEEDE) model} and developed a two-step approach accordingly to mitigate the impact of orbital errors. In a carrier-phase-based positioning system with ORBCOMM satellites, additional system parameters were introduced into the carrier-phase observation model to account for the influence of system errors~\cite{xie2024carrier}. Additionally, a double-difference Doppler formulation has been used to mitigate receiver and satellite clock errors and partially remove orbit biases~\cite{hasan2024double}. Notably, each of these methods provides a deterministic compensation approach for specific observables, while the statistical prior of orbital errors has not been exploited. 
        
    Another category of approaches leverages machine learning techniques. Recent advances have led to the application of machine learning to \ac{leo} orbit calibration through two primary paradigms. One is machine learning-based offline orbit prediction: By training on historical orbit data, the model compensates for errors in orbital dynamics prediction, thereby offering more accurate orbital information~\cite{caldas2024precise, kassas2024leo}. The other is the integration of machine learning into navigation filters for online orbit/position estimation~\cite{mortlock2021assessing}. Overall, these learning-based methods provide flexible data-driven tools for reducing orbit prediction errors. Nonetheless, their performance may also be limited by generalization capability and depends on the representativeness of the training data, the satellite regime, and the consistency between training and deployment conditions.
    
\subsection{Main Contributions}
    Motivated by the aforementioned research gaps, this work aims to establish a systematic understanding of the impact of orbital errors on \ac{leo}-based positioning, characterize their statistical behavior, and develop orbit calibration and \ac{ue} positioning strategies that leverage the learned statistical prior. To this end, we first establish an orbit model and the associated orbit propagators, followed by a statistical characterization of orbital errors using historical \ac{tle} data collected from several \ac{leo} constellations. Based on the established orbit model, we derive the \ac{mcrb}, which serves as a theoretical positioning error bound and offers fundamental insights into how orbital errors affect positioning performance. Building on the learned orbital error statistics, we further develop a Bayesian orbit calibration framework that explicitly incorporates these statistics as prior information to improve calibration accuracy. Although motivated by LEO positioning, the proposed orbit model, statistical orbital error characterization, MCRB-based analysis, and Bayesian orbit calibration framework are formulated in a general manner and can be applied to other satellite systems like \ac{gnss}. Finally, we present a case study, including terrestrial \ac{bs}-assisted orbit calibration and the subsequent \ac{ue} positioning based on the calibrated orbit, to demonstrate the effectiveness of the proposed framework. 
    
    The main contributions of this paper are summarized as follows:
    \begin{itemize}
        \item We establish an age-dependent statistical model to characterize the orbital errors of an orbit propagator given an orbital information dataset. Based on this model, we evaluate the statistical behavior of orbit errors in multiple \ac{leo} constellations using their historical \ac{tle} data, revealing how orbital mismatch evolves with the orbital information age and varies across constellations.
        
        \item We derive the \ac{mcrb} as a theoretical positioning error bound under orbit mismatch, providing a quantitative measure of the achievable positioning accuracy under orbital errors. The analysis offers valuable insights into how the evolution of orbital errors affects positioning performance. Further, we investigate how orbital errors propagate to different geometric observables by performing sensitivity analyses with respect to individual orbital elements, thereby uncovering the dominant factors behind the resulting positioning degradation.
        
        \item We develop a Bayesian orbit calibration framework that incorporates the learned orbital error statistics as prior information. Given observations from a few known calibration anchors, the resulting calibration problem is formulated as a \ac{map} estimation problem, which enables statistically regularized orbit correction under mismatched orbital information.
        
        \item We conduct a representative case study to validate the proposed Bayesian orbit calibration framework. Particularly, to demonstrate its practical impact on final positioning performance, we further develop a low-complexity \ac{ue} position estimator, enabling an end-to-end evaluation of the complete calibration-then-positioning pipeline. The results verify that the proposed Bayesian framework yields considerably more accurate orbit calibration compared to state-of-the-art benchmarks, which subsequently translates into improved \ac{ue} positioning accuracy.
    \end{itemize}
    
    This paper is organized as follows. Section~\ref{sec_orbit_model} presents the orbit model, including the adopted coordinate frames and the satellite orbit formulation. Section~\ref{sec_orb_err_modeling_and_ana} develops the statistical orbital error model and presents a set of evaluations using historical \ac{leo} \ac{tle} data. Section~\ref{sec_impact_of_orbital} investigates the impact of orbital errors on positioning performance by analyzing the \ac{mcrb} under orbital mismatch. Section~\ref{sec_orb_calibration} proposes the Bayesian orbit calibration framework based on the learned statistical orbital error model. A case study is presented in Section~\ref{sec_CS} to validate the proposed orbit calibration framework, followed by the corresponding simulation results in Section~\ref{sec_simumation}. Finally, Section~\ref{sec_conclusion} concludes the paper.

    Throughout this paper, we adopt the following notational conventions. Vectors are denoted by bold lowercase (e.g.,~\(\mathbf{x}\)), and matrices by bold uppercase (e.g.,~\(\mathbf{X}\)). All vectors are column vectors. The~\(n\)-th element of~\(\xv\) is written as~\([\xv]_n\), and the~\((m,n)\) entry of~\(\Xm\) as~\([\Xm]_{m,n}\). We use~\([\Xm]_{m:n,k:\ell}\) to denote the submatrix comprising rows~\(m\) through~\(n\) and columns~\(k\) through~\(\ell\). We use \(\mathrm{diag}(\cdot)\) as the diagonal operator. Let \(\Xm \times \Ym\), \(\Xm \otimes\Ym\), and \(\Xm \cdot \Ym\) denote the cross-product, the Kronecker product, and the dot product between \(\Xm\) and \(\Ym\), respectively. The Euclidean norm of a vector~\(\xv\) is denoted by~\(\|\xv\|_2\). The functions are denoted by script letters (e.g.,\(~\mathscr{F}\)). Finally,~\(\mathbf{1}_L\) denotes the~\(L\)-dimensional all-ones column vector, and \(\Id_L\) denotes the \(L\)-by-\(L\) identity matrix. 
    
\section{Coordinate Geometry and Orbit Model}\label{sec_orbit_model}
   In this section, we introduce the geometric framework required for the subsequent analysis. We first define the coordinate frames adopted throughout this paper and then introduce the orbital elements used to describe satellite motion.
   
\subsection{Coordinate Frames}\label{subsec_coordinate_frames}
\begin{figure*}
        \centering
        \includegraphics[width=0.95\linewidth]{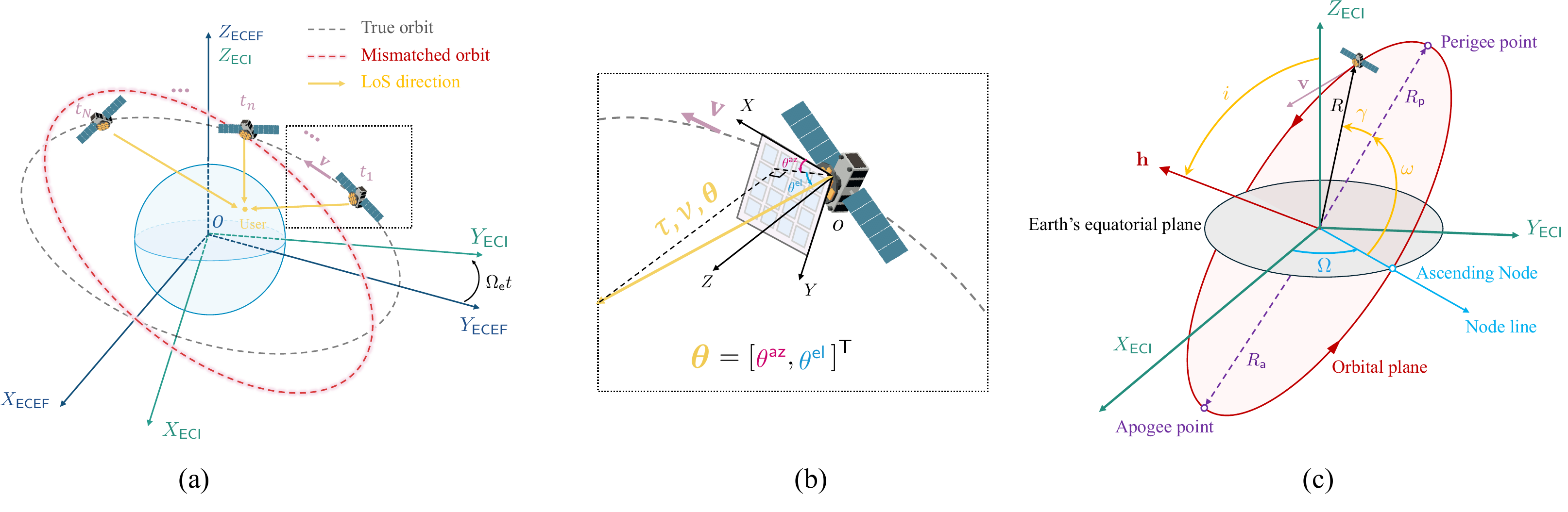}
        \vspace{-1.5em}
        \caption{
		Illustration of the \ac{leo}-based positioning scenario. (a) Relationship between the ECI and ECEF frames. The two frames share the same Z-axis and a common origin at Earth's center. (b) Demonstration of the \ac{sl} frame. (c) Illustration of orbital elements in the ECI frame.}
        \label{fig_sysDemo}
    \end{figure*}

    We define the following four coordinate systems: the \ac{eci} frame, the \ac{ecef} frame, the \ac{sl} frame, and the \ac{enu} local frame. The relationship among them is illustrated in Fig.~\ref{fig_sysDemo}(a) and~\ref{fig_sysDemo}(b). 
    \subsubsection{Earth Related Coordinate Frames}
    The \ac{eci} frame is a right-handed Cartesian coordinate frame centered on the center of the Earth, with its axes fixed relative to distant stars. The~\(Z_\mathsf{ECI}\) axis aligns with the Earth's rotational axis, and the~\(X_\mathsf{ECI}\) axis points toward the vernal equinox. The \ac{eci} frame is commonly treated as an inertial reference frame for satellite dynamics and orbital mechanics.
    
    The \ac{ecef} frame is Earth-fixed, and it shares the origin and~\(Z\) axis with the \ac{eci} but co-rotates with Earth about the~\(Z\) axis. It is a non-inertial frame and is typically used to describe the positions of ground-based users and Earth-fixed objects at a given time. The transformation from the \ac{eci} frame to the \ac{ecef} frame can be performed by a rotation about the~\(Z\)-axis with the Earth's rotation angle. Specifically, it is given by
    \begin{equation} \label{eq_ECI_to_ECEF}
    \pv_{\mathsf{ECEF},t} = \Rm_{\mathsf{ECI} \rightarrow \mathsf{ECEF}} \cdot \mathbf{p}_{\mathsf{ECI},t},
    \end{equation}
    where 
    \begin{equation} 
        \Rm_{\mathsf{ECI} \rightarrow \mathsf{ECEF}}
        =
        \begin{bmatrix}
        \cos(\Omega_e t) & -\sin(\Omega_e t) & 0 \\
        \sin(\Omega_e t) & \cos(\Omega_e t) & 0 \\
        0 & 0 & 1
        \end{bmatrix}.
    \end{equation}
    Here, \(\pv_{\mathsf{ECI},t}\) and \(\pv_{\mathsf{ECEF},t}\) are the \ac{ue} positions at time \(t\) in the ECI and ECEF frames, respectively. The parameter~\( \Omega_e \) is the angular velocity of the Earth's rotation, and~\( t \) is the elapsed time since the \emph{Greenwich Mean Sidereal Time (GMST)} reference epoch at J2000.0.
    
    \subsubsection{Local Coordinate Frames}
    The \ac{sl} frame is a right-handed Cartesian frame attached to the satellite body. As plotted in Fig.~\ref{fig_sysDemo}(b), the~\(Z\)-axis points towards the center of Earth, while the~\(X\)-axis is aligned with the velocity direction. \footnote{Equivalently, the orientation of the SL frame with respect to the ECEF frame can be parameterized using the conventional pitch-roll-yaw representation.} This frame is used to describe antenna orientations and channel observables relative to satellites. Given the satellite position vector~\( \pv_{\mathsf{s}} \in \mathbb{R}^{3} \) and velocity vector~\( \vv \in \mathbb{R}^{3} \) in the \ac{ecef} frame, the \ac{sl} frame axes can be computed (in the \ac{ecef} frame) as:
    \begin{align}\label{eq_rotMat_xyz}
    \zv = -\frac{\pv_{\mathsf{s}}}{\|\pv_{\mathsf{s}}\|_2}, \quad
    \xv = \frac{\vv}{\|\vv\|_2}, \quad
    \yv = \frac{\zv \times \xv}{\|\zv \times \xv\|_2}.
    \end{align}
    The rotation matrix from the \ac{sl} frame to the \ac{ecef} frame is then given by
    \begin{equation} \label{eq_rotMat}
    \Rm_{\mathsf{SL}\to\mathsf{ECEF}}= 
    \begin{bmatrix}
    \xv \quad \yv \quad \zv
    \end{bmatrix}
    \in \mathbb{R}^{3 \times 3}.
    \end{equation} 

    For ground objects (e.g., \acp{ue} and \acp{bs}), we define a local \ac{enu} frame, whose axes point to east, north, and upward (radially outward from Earth's center), respectively. The rotation from \ac{enu} to \ac{ecef} at geodetic latitude \(\phi_{\mathsf E}\) and longitude \(\lambda_{\mathsf E}\) is given by
    \begin{equation}
    \pv_{\mathsf{ECEF}}=\Rm_{\mathsf{ENU}\to\mathsf{ECEF}} \cdot \pv_{\mathsf{ENU}}, 
    \end{equation}
    where
    \begin{equation}
    \Rm_{\mathsf{ENU}\to\mathsf{ECEF}} \!=   \!\!
    \begin{bmatrix} 
    -\sin\lambda_\mathsf{E} &    -\sin\phi_\mathsf{E}\cos\lambda_\mathsf{E} &   \cos\phi_\mathsf{E}\cos\lambda_\mathsf{E}  \\
    \cos\lambda_\mathsf{E} &    -\sin\phi_\mathsf{E}\sin\lambda_\mathsf{E} &    \cos\phi_\mathsf{E}\sin\lambda_\mathsf{E}  \\
    0 &    \cos\phi_\mathsf{E} &   \sin\phi_\mathsf{E}
    \end{bmatrix}  .
    \label{eq_Rn2e}
    \end{equation}

    \subsection{Orbit Model}\label{subsec_orbit_model}
    
    \subsubsection{Orbital Elements}
    The orbit of a satellite is generally described in the \ac{eci} frame and can be parameterized by the classical Keplerian elements collected in a vector \(\ov\in\mathbb{R}^{6}\) as
    \begin{equation}\label{eq_def_orb}
        \ov \triangleq [a,e,i,\Omega, \omega, \gamma]^\TT.
    \end{equation}
    As shown in Fig.~\ref{fig_sysDemo}(c), \(a\) is the semi-major axis defined as the mean value of the apogee distance~\(R_\mathsf{a}\) and the perigee distance~\(R_\mathsf{p}\), calculated as~\(a = (R_\mathsf{a}+R_\mathsf{p})/2\); \(e\) is the eccentricity, defined as~\(e = {(R_\mathsf{a}-R_\mathsf{p})}/{(R_\mathsf{a}+R_\mathsf{p})}\); \(i\) is the inclination, defined as the angle between the satellite's specific angular momentum vector~\(\hv\triangleq \pv_\mathsf{s}\times \vv\) and~\(Z_\mathsf{ECI}\);~\(\Omega\)~is the \ac{raan} measured from the inertial reference direction to the ascending node; \(\omega\) is the argument of periapsis, defined as the angle between the ascending node and the periapsis point; \(\gamma\) is the true anomaly, defined as the angle between the periapsis point and the satellite's current position. 

    It is trivial to see that the first five orbital elements \(\{a,e,i,\Omega,\omega\}\) determine the shape and orientation of the orbit, whereas the true anomaly \(\gamma\) specifies the satellite's instantaneous location along the orbit. Thus, when the orbit is fixed, the satellite position can be uniquely determined by~\(\gamma\). In practice, however, orbital perturbations cause all six orbital elements to evolve over time. Consequently, to determine the instantaneous satellite position, the complete orbital element vector \(\ov\) is required.

    \subsubsection{Orbit Evolution and Prediction}
    Typically, the orbital elements in \(\ov\) can be obtained from publicly released \ac{tle} data~\cite{KelsoGPFormats}. However, since these data are not updated in real time, the orbital element vector at intermediate epochs must be predicted. Specifically, given an orbital element vector $\ov_{t_0} \triangleq [a_{t_0},e_{t_0},i_{t_0},\Omega_{t_0}, \omega_{t_0}, \gamma_{t_0}]^\TT$ at time $t_0$, the corresponding orbital element vector at any future time $t>t_0$, denoted by~\(\ov_t\), can be predicted according to the underlying orbital dynamics. This prediction task can be handled by a variety of orbit propagators, including the standard SGP4~\cite{Vallado2006SGP4}, the semi-analytical propagators based on the Draper Semi-analytical Satellite Theory (DSST)~\cite{danielson1995semianalytic}, and high-fidelity numerical propagators such as the High-Precision Orbit Propagator (HPOP)~\cite{STK_HPOP_Help}, etc. To make the paper self-contained, we present a simplified orbit propagator recommended by the QuaDRiGa platform~\cite{Jaeckel2021} in Appendix~\ref{sec_app_orb_model}. In practice, however, SGP4 is arguably the most widely used in the satellite community. 
    
    \subsubsection{Satellite Position Determination}
    Given $\ov_t$, the satellite's distance from the Earth's center and its position in the \ac{eci} frame are immediately determined as:
    \begin{equation}
        R_t = \frac{a_t (1 - e_t^2)}{1 + e_t \cos (\gamma_t)},
    \end{equation}
    \begin{equation} \label{eq_Ps}
        \hspace{-0.5em}\mathbf{p}_{\mathsf{ECI},t}  \!=  \!R_t  \!\!
    \begin{bmatrix}\!
     \cos(\omega_t  \!+\!   \gamma_t) \cos\Omega_t  \!-\!  \sin(\omega_t  \!+\!   \gamma_t) \sin\Omega_t \cos i_t \!\\
     \!\cos(\omega_t   \!+\!   \gamma_t) \sin\Omega_t  \!+\!  \sin(\omega_t   \!+\!   \gamma_t) \cos\Omega_t \cos i_t \\
    \sin(\omega_t  \!+\!  \gamma_t) \sin i_t\!
    \end{bmatrix} \! .
    \end{equation}

\section{Orbital Error Modeling and Analysis}\label{sec_orb_err_modeling_and_ana}
Based on the orbit model presented above, this section models, evaluates, and analyzes \ac{leo} orbital errors.

\subsection{An Age-Dependent Orbital Error Model}\label{subsec_orbital_error_statistics}
While the orbit propagators mentioned in Section~\ref{subsec_orbit_model} provide a tractable means of predicting the orbital evolution at arbitrary times, orbital prediction errors may arise. This is because the true evolution of satellite motion is influenced by additional factors, such as higher-order gravitational harmonics, atmospheric drag, solar radiation pressure, and third-body perturbations from the Sun and the Moon, which cannot be perfectly captured. As a result, the resulting orbital mismatch inevitably accumulates over time.

We can quantify the orbital prediction errors associated with the propagator by comparing its predictions with ground-truth orbital elements. Although the true orbital elements are generally not available, observed orbital information such as historical \ac{tle} data serves as a surrogate. Specifically, let's denote the orbit prediction using a propagator by 
\begin{equation}
    \tilde{\ov}_{t} = \mathscr{F}(\ov_{t-T}, T),
\end{equation}
where $\tilde{\ov}_{t}$ denotes the predicted orbital element vector at time $t$, $\mathscr{F}(\ov_{t-T}, T)$ denotes the propagator that predicts $\tilde{\ov}_{t}$ given $\ov_{t-T}$, and thus $T$ is the age of orbital information. For a given satellite, assume that there is a set of observed orbital element vectors $\{\ov_t\}_{t=t_1}^{t_N}$ sampled at epochs in $\mathcal{S}=\{t_1,t_2,\dots,t_N\}$. Then, one can compute the sample mean of the prediction errors as 
\begin{equation}\label{eq_mu_o}
    \muv_\ov(T) =  \frac{1}{|\mathcal{S}_{T}|}
    \sum_{(t_i,t_j)\in\mathcal{S}_{T}}
     (\mathscr{F}(\ov_{t_i}, t_j-t_i) - \ov_{t_j}),
\end{equation}
where $\mathcal{S}_{T}\triangleq\{(t_i,t_j)\,|\,t_i,t_j\in\mathcal{S},\,|t_j-t_i-T|<\varepsilon\}$ with $\varepsilon$ denoting a small tolerance. Further, the sample covariance matrix can be computed as 
\begin{equation}\label{eq_sigma_o}
\Sigmam_{\ov }(T)
\!=\!
    \frac{1}{|\mathcal{S}_{T}|}   
    \!\!\sum_{(t_i,t_j)\in\mathcal{S}_{T}}  \!\!\!
    \big(
    \deltav_{i,j}
    -
    \muv_{\ov }(T)
    \big)
    \big(
    \deltav_{i,j}
    -
    \muv_{\ov }(T)
    \big)^\TT\!,
\end{equation}
where $\deltav_{i,j} \triangleq \mathscr{F}(\ov_{t_i}, t_j-t_i) - \ov_{t_j}$. 

Following the Gaussian uncertainty modeling commonly adopted in orbit uncertainty characterization and covariance realism analysis~\cite{cano2022improving,yanez2019gaussianity}, we approximate the orbital prediction errors by a Gaussian random vector. Using the sample mean and covariance matrix learned from data, the orbital error vector is characterized by a Gaussian random vector as
\begin{equation} \label{eq_Delta_o}
    \Delta \ov (T)
    \sim
    \mathcal{N} \big(
    \muv_{\ov }(T),
    \Sigmam_{\ov }(T)
    \big),
\end{equation}
where \(\Delta \ov(T) \triangleq \mathscr{F}(\ov_{t-T}, T)-\ov_{t}\) denotes the orbital element prediction error of the orbit propagator $\mathscr{F}$ after an evolution interval of $T$. Note that the distribution of $\Delta \ov(T)$ is assumed to depend only on the propagation interval and is invariant with respect to the initial epoch.

\subsection{Evaluation and Analysis Based on LEO TLE Data}
Using the developed model, this subsection presents an evaluation of the orbital error statistics for five exemplary \ac{leo} satellite constellations. 
The reference orbital element vectors $\{\ov_t\}_{t=t_1}^{t_N}$ are derived from historical \ac{tle} data collected from \emph{space-track.org}. For each constellation, we collect the records of all satellites in the selected launch group that were still operational until 17/03/2026. 

\begin{table}[t]
    \centering
    \caption{Constellations Used for the Orbital Error Evaluation}
    \label{tab_satellite_cons_info}
    \begin{tabular}{lccccc}
         \toprule
         Constellation & COSPAR ID &  Alt. [km] & \# Sats & \# TLE & Launch \\
        \midrule
        Starlink-187 &2024-187& 360 & 7  & 7137  & 2024 \\
        Starlink-082 &2021-082& 576 & 6  & 22717 & 2021 \\
        Kuiper    &2025-088& 630 & 20 & 14794 & 2025 \\
        ORBCOMM   &2015-081& 700 & 5  & 31075 & 2015 \\
        Iridium   &2017-003& 780 & 5  & 34875 & 2017 \\
        \bottomrule
    \end{tabular}
\end{table}
 \begin{figure}
    \centering
\definecolor{mycolor1}{rgb}
{0.78039,0.13333,0.15686}%
\definecolor{mycolor2}{rgb}{0.96078,0.52549,0.49804}%
\definecolor{mycolor3}{rgb}{0.97647,0.56078,0.20392}%
\definecolor{mycolor4}{rgb}{0.48235,0.58431,0.77647}%
\definecolor{mycolor5}{rgb}{0.07843,0.31765,0.48627} %
\definecolor{mycolor6}{rgb}{0.41961,0.59608,0.76863}%
\definecolor{mycolor7}{rgb}{0.41569,0.68627,0.17647}%
\definecolor{mycolor8}{rgb}{0.96078,0.52549,0.49804}%
\begin{tikzpicture}
\begin{axis}[%
width=2.7in,
height=1.8in,
scale only axis=true,
enlarge x limits=false,
enlarge y limits=false,
xmin=0,
xmax=24.1,
xlabel={TLE Update Interval $T$ [hr]},
xlabel style={font=\footnotesize, yshift=2pt}, 
xtick = {1,3,5,7,9,11,13,15,17,19,21,23},
ylabel={Number of Update Intervals},
ylabel style={font=\footnotesize, yshift=-5pt}, 
ymode=linear,
ymin=3,
ymax=17000,
yminorticks=true,
tick label style={font=\footnotesize},                  
axis background/.style={fill=white},                    
axis lines=box,                                         
grid style={dotted, draw=gray!50},                      
xmajorgrids,
xminorgrids,
ymajorgrids,
yminorgrids,
legend style={
    at={(1,1.04)},                                   
    anchor=south east,
    legend columns=3,
    draw=black,                                       
    font=\footnotesize,                                
    /tikz/every even column/.append style={column sep=0.5em},
    nodes={scale=0.9, transform shape}                 
},
]
\addplot[ybar stacked, fill=mycolor1, draw=black, area legend] table[row sep=crcr] {%
1	2974\\
3	2083\\
5	2699\\
7	2854\\
9	3253\\
11	8203\\
13	3817\\
15	1844\\
17	1604\\
19	1309\\
21	1650\\
23	1265\\
};
\addplot[forget plot, color=white!15!black] table[row sep=crcr] {%
0	0\\
24	0\\
};
\addlegendentry{Iridium}

\addplot[ybar stacked, fill=mycolor2, draw=black, area legend] table[row sep=crcr] {%
1	2532\\
3	1610\\
5	1846\\
7	3542\\
9	5674\\
11	839\\
13	1832\\
15	2391\\
17	2678\\
19	2575\\
21	1077\\
23	2245\\
};
\addplot[forget plot, color=white!15!black] table[row sep=crcr] {%
0	0\\
24	0\\
};
\addlegendentry{ORBCOMM}

\addplot[ybar stacked, fill=mycolor3, draw=black, area legend] table[row sep=crcr] {%
1	2006\\
3	1153\\
5	1158\\
7	3033\\
9	1958\\
11	731\\
13	687\\
15	769\\
17	1170\\
19	354\\
21	377\\
23	500\\
};
\addplot[forget plot, color=white!15!black] table[row sep=crcr] {%
0	0\\
24	0\\
};
\addlegendentry{Kuiper}

\addplot[ybar stacked, fill=mycolor4, draw=black, area legend] table[row sep=crcr] {%
1	848\\
3	687\\
5	457\\
7	1222\\
9	191\\
11	561\\
13	634\\
15	143\\
17	444\\
19	550\\
21	302\\
23	497\\
};
\addplot[forget plot, color=white!15!black] table[row sep=crcr] {%
0	0\\
24	0\\
};
\addlegendentry{Starlink-187}

\addplot[ybar stacked, fill=mycolor5, draw=black, area legend] table[row sep=crcr] {%
1	2276\\
3	1820\\
5	1444\\
7	4136\\
9	4154\\
11	814\\
13	989\\
15	1233\\
17	3021\\
19	632\\
21	566\\
23	456\\
};
\addplot[forget plot, color=white!15!black] table[row sep=crcr] {%
0	0\\
24	0\\
};
\addlegendentry{Starlink-082}

\node[
    font=\footnotesize,
    fill=white,
    draw=none,
    align=left,
    inner sep=2pt,
    anchor=north east
] at (axis description cs:0.98,0.98) { $T>12$ hr:\\
\textcolor{mycolor5}{- Starlink-082: 35.5\%}\\
\textcolor{mycolor4}{- Starlink-187: 44.4\%}\\
\textcolor{mycolor3}{- Kuiper: 31.8\%}\\
\textcolor{mycolor2}{- ORBCOMM: 48.4\%}\\
\textcolor{mycolor1}{- Iridium: 36.7\%}
};
\end{axis}
\end{tikzpicture}%
    \vspace{-1em}
    \caption{Distribution of orbital information update intervals (i.e., orbital information ages) for different LEO constellations in the \ac{tle} dataset.}
    \label{fig_orberr_distribution}
\end{figure}

Notably, satellites in this dataset are not all tracked and updated at the same frequency. To illustrate this, we first examine the \ac{tle} update intervals for different \ac{leo} constellations. As shown in Fig.~\ref{fig_orberr_distribution}, the update intervals vary significantly across and within constellations. In particular, a non-negligible fraction of \ac{tle} samples falls into update intervals longer than 12 hours, indicating that stale orbital information is not merely an occasional outlier but a common and practical concern. 
 We then evaluate the orbital error statistics for each available update interval \(T\) using the SGP4 propagator as the orbit propagator \(\mathscr{F}\). Specifically, for each \(T\), the corresponding orbital prediction error samples are computed to obtain the mean vector \(\muv_{\ov}(T)\) and covariance matrix \(\Sigmam_{\ov}(T)\) according to \eqref{eq_mu_o} and \eqref{eq_sigma_o}. As shown in Fig.~\ref{fig_orberr_cov}, the standard deviations of the orbital element errors differ across constellations and depend on the orbital information update interval. In general, as the update interval increases, the orbital errors become more pronounced, and their spread becomes larger. However, the effect of the update interval is not identical across constellations. 
 These effects are influenced by both constellation maturity and orbital altitude. For newly deployed constellations, such as Starlink-187 launched in 2024 and Kuiper launched in 2025, the satellites may still operate under an unstable steady-state condition. In such cases, orbit adjustment and constellation reconfiguration can lead to stronger temporal variations in the orbital errors. Thus, timely updates are particularly important for such constellations. In addition, satellites at lower orbital altitudes are generally more sensitive to atmospheric drag~\cite{KUANG2014243}. Therefore, compared with higher-altitude constellations such as Iridium, lower-altitude constellations are more susceptible to the growth of orbital mismatch when the orbital information becomes stale.

\begin{figure}
    \centering
    \input{fig_revise1/fig_orberr_cov}
    \vspace{-2em}
    \caption{Standard deviations of the six orbital elements for different constellations under different orbital information update intervals.}
    \label{fig_orberr_cov}
\end{figure}

\section{Impact of Orbital Errors on Positioning}\label{sec_impact_of_orbital}
Building on the developed statistical characterization of \ac{leo} orbital errors, this section evaluates their impact on terrestrial \ac{ue} positioning through the \ac{mcrb} theory.

\subsection{Single-Orbit Positioning}\label{subsec_single_orb_pos}

To gain insight into the fundamental effect of orbital errors on positioning performance, we begin by focusing on a single-orbit scenario. Specifically, we consider a positioning system consisting of a single \ac{leo} satellite and serving a static terrestrial \ac{ue}. As the satellite traverses its orbit, it transmits downlink signals to the \ac{ue} at multiple time epochs \(\ell\in\{1,2,\dots, L\}\) from different locations along the orbit. The \ac{ue} estimates its location based on these received signals. This estimation relies on knowledge of the satellite states at different epochs, which is determined based on the orbital element vector as shown in~\eqref{eq_Ps}. Since the duration of these $L$ signal transmissions is typically much shorter than the time scale over which the orbital error evolves, the orbital elements are assumed to remain constant throughout the $L$-epoch observation window, i.e., the time interval during which the considered satellite is visible.

\begin{remark}\label{rmk1}
\emph{Throughout this paper, we distinguish between two timescales. The \emph{slow-time}, on the order of hours, captures the evolution of the satellite orbit, which is used for \ac{tle} updates and orbit propagation. In contrast, the \emph{fast-time}, on the order of seconds, refers to the observation epochs at which signals are transmitted for \ac{ue} positioning or orbit calibration (as discussed in Section~\ref{sec_orb_calibration}), during which the satellite orbit is assumed to remain unchanged.}
\end{remark}

\begin{remark}\label{rmk2}
\emph{Accordingly, we assume that the $L$ signal transmissions occur at slow-time epoch $t$ with true orbit $\ov_t$. The available orbital information, however, is predicted based on a previous orbital element vector $\ov_{t_0}$ at time $t_0<t$. While $\ov_{t_0}$ is assumed to be accurate, the propagated orbit may deviate from the true orbit.}
\end{remark}

To estimate the \ac{ue} position, geometric observables (e.g., propagation delays, carrier phases, \acp{aod}, and Doppler-related quantities) usually need to be extracted from the received signals first. The extraction of geometric observables, however, depends strongly on the underlying system architecture, operating frequency, and signal waveform. To provide a general framework, here we simply denote the obtained geometric observables at time epoch $\ell$ as $\hat{\etav}_\ell=\etav_\ell+\uv_\ell$, where $\etav_\ell$ denotes the ground-truth observables and \(\uv_\ell\sim\mathcal N(\mathbf 0,\Sigmam_\ell)\) denotes a zero-mean Gaussian estimation error. A concrete example of these geometric observables is provided in Section~\ref{subsec_pos_performance}. Concatenating the observables over $L$ time epochs, we have
\begin{equation}
    \hat{\etav}=\etav+\uv,
\end{equation}
where $\hat{\etav}=[\hat{\etav}_1^\TT,\dots,\hat{\etav}_L^\TT]^\TT$, ${\etav}=[{\etav}_1^\TT,\dots,{\etav}_L^\TT]^\TT$, and $\uv\sim\mathcal{N}(\mathbf{0},\Sigmam_{\etav})$ with $\Sigmam_{\etav}=\mathrm{blkdiag}\{\Sigmam_1,\dots,\Sigmam_L\}$. Therefore, the positioning problem refers to estimating the \ac{ue} position based on $\hat{\etav}$. 

Nonetheless, the forward model relating the observable vector $\etav$ to the \ac{ue} position may involve additional unknown parameters. For example, when propagation delay observables are employed in an asynchronous system, the clock bias between the \ac{ue} and satellite must be treated as an additional unknown~\cite{Zheng2024JrCUP}. Likewise, when carrier phase observables are utilized, an integer ambiguity vector is further introduced into the estimation problem~\cite{Zheng020235G}. To accommodate these scenarios in a unified manner, we define the \ac{ue} state vector as
\begin{equation}\label{eq_ue_state}
    \xiv \triangleq \big[\pv^\TT,\rv^\TT\big]^\TT,
\end{equation}
where $\pv\in\mathbb{R}^3$ denotes the unknown \ac{ue} position in the \ac{ecef} frame, and $\rv$ collects all other unknown parameters. The forward model can then be expressed as $\etav=\mathscr{G}(\xiv\,|\,\ov_t)$, where the orbital element vector $\ov_t$ is treated as a set of known model parameters. 

Since the ground-truth $\ov_t$ is unavailable, the model that generates the true observables differs from the model used for position estimation. Accordingly, the observation model and the estimation model can be written as follows:
\begin{align}
   \mathsf{Observation\ Model:\ } \hat{\etav} &= \mathscr{G}(\xiv\,|\,\ov_t) + \uv,\\
   \mathsf{Estimation\ Model:\ } \hat{\etav} &= \mathscr{G}\big(\xiv\,|\,\mathscr{F}(\ov_{t_0}, t-t_0)\big) + \uv.
\end{align}
The model difference, i.e., the orbital error $\mathscr{F}(\ov_{t_0}, t-t_0)-\ov_t$, is modeled as a Gaussian random vector as demonstrated in~\eqref{eq_Delta_o}, whose strength depends on the orbital information age $t-t_0$. For brevity, we denote $\mathscr{F}(\ov_{t_0}, t-t_0)$ by $\tilde{\ov}_t$. The maximum likelihood estimation of $\xiv$ can therefore be formulated as
\begin{equation}\label{eq_pos_only}
    \hat{\xiv}
    =
    \arg\min_{\xiv}\ 
    \bigl(
    \hat{\etav}
    -
    \mathscr{G}(\xiv|\tilde{\ov}_t)
    \bigr)^\TT
    \Sigmam_{\etav}^{-1}
    \bigl(
    \hat{\etav}
    -
    \mathscr{G}(\xiv|\tilde{\ov}_t)
    \bigr).
\end{equation}
In the following, we derive the performance bound of this maximum likelihood estimation problem.

\subsection{MCRB Derivation}
Since the \ac{ue} adopts the mismatched orbital information \(\tilde{\ov}_t\) instead of the true orbital elements \(\ov_t\), the resulting positioning model is mismatched. Therefore, we calculate and analyze the \ac{mcrb}, which provides a lower bound on the estimation error under such model mismatch~\cite{fortunati2017performance}. In what follows, we omit the time subscript \(t\) in \(\ov_t\) and \(\tilde{\ov}_t\) for notational simplicity. 

Let \(f_\mathsf{T}\) and \(f_\mathsf{M}\) denote the true and mismatched likelihood functions, respectively. Under the Gaussian estimation error assumption, the log-likelihoods associated with the true observation model and the mismatched estimation model are
    \begin{align}
        \ln   f_\mathsf{T}  =  -\frac{1}{2}   (  \hat{\etav}   -  \mathscr{G} ( \xiv  |  \ov )  )^\TT   \Sigmam_{\etav}^{-1}        
        (  \hat{\etav}  - \mathscr{G} ( \xiv  |  \ov )   ), \\
        \ln   f_\mathsf{M}  =  -\frac{1}{2}   (  \hat{\etav}   -  \mathscr{G} ( \xiv  |  \tilde{\ov} )  )^\TT   \Sigmam_{\etav}^{-1}  (  \hat{\etav}  - \mathscr{G} ( \xiv  |  \tilde{\ov} )   ).
        \end{align}
        
        When estimation is performed under the mismatched model \(f_{\mathsf M}\), the \ac{mcrb} associated with a pseudo-true parameter \(\xiv_0\) is given by
        \begin{equation}
           \text{MCRB}(\xiv_0)  \triangleq \Am_{\xiv_0}^{-1}  \Bm_{\xiv_0}   \Am_{\xiv_0}^{-1}.
        \end{equation}

        Then, the \ac{mse} admits the following lower bound~\cite{fortunati2017performance}:
    \begin{equation}\label{eq_lbm}
        \text{LBM}(\xiv_0, \bar{\xiv})  \triangleq  \underbrace{\Am_{\xiv_0}^{-1}  \Bm_{\xiv_0}   \Am_{\xiv_0}^{-1} }_{\text{MCRB}(\xiv_0)}  + \underbrace{(\bar{\xiv}  -  \xiv_0)(\bar{\xiv}  -  \xiv_0)^\TT}_ {\text{Bias}(\xiv_0)}.
    \end{equation}
    Here, \(\bar{\xiv}\) is the true unknown parameter vector, \(\xiv_0\) is the \emph{pseudo-true} parameter that minimizes the Kullback-Leibler divergence (KLD) between the true and the mismatched models, and \(\Am_{\xiv_0}\) and \(\Bm_{\xiv_0}\) are two generalized \acp{fim}. The pseudo-true parameter \(\xiv_0\) and the generalized information matrices \(\Am_{\xiv_0}\) and \(\Bm_{\xiv_0}\) are given by~\cite{fortunati2017performance}
    \begin{align}
     &\xiv_0= \arg \min_{\xiv}\ 
         \mathrm{KLD}\bigl(f_\mathsf{T}(\hat\etav |\bar\xiv)
               \|  f_\mathsf{M}(\hat\etav   |\xiv)\bigr) 
      \notag\\
&\quad =  \arg\min_{\xiv}  \tfrac12 
         \bigl(\mathscr{G}(\bar\xiv  |  \ov )
              \!-\!   \mathscr{G}(\xiv  |  \tilde\ov ) \bigr)^{\TT}
          \Sigmam_{\etav}^{-1}
           \bigl(\mathscr{G}(\bar\xiv  |  \ov )
              \!-\! \mathscr{G}(\xiv  |  \tilde\ov ) \bigr),\notag \\
&\big[\Am_{\xiv_0}\big]_{i,j} 
        =   \left(   \frac{\partial^2 \mathscr{G}\big(\xiv  |         \tilde{\ov} \big)}{\partial {[{\xiv]_i}} \partial {[\xiv]_j}}   \right)^\TT   \Sigmam_{\etav}^{-1}   \left(\etav   -  \mathscr{G}\left(\xiv  |      \tilde{\ov}     \right)\right) \bigg|_{\xiv=\xiv_0}\notag\\
        &\qquad\qquad\qquad -  \left(  \frac{\partial \mathscr{G}\left(\xiv|    \tilde{\ov}   \right)}{\partial {{[\xiv]_i}}}   \right)^\TT     \Sigmam_{\etav}^{-1}   \left( \frac{\partial \mathscr{G}\left(\xiv| \tilde{\ov} \right)}{\partial {[\xiv]_j}} \right)  \bigg|_{\xiv=\xiv_0},\notag\\
&\big[\Bm_{\xiv_0}\big]_{i,j}  
        =    \bigg(  \! \frac{\partial \mathscr{G} \left(\xiv  |      \tilde{\ov}   \right)}{\partial {{[\xiv]_i}}}   \!\bigg)^\TT   \!\!  \Sigmam_{\etav}^{-1}  \tilde{\Sigmam}_{\etav}(\xiv)\Sigmam_{\etav}^{-1}  \! \bigg(\!  \frac{\partial \mathscr{G}\left(\xiv  |     \tilde{\ov} \right)}{\partial {[\xiv]_j}} \! \bigg)  \!\bigg|_{\xiv=\xiv_0}\!,\notag
\end{align}
    where~\(\tilde{\Sigmam}_{\etav}(\xiv)  =  \Sigmam_{\etav} + \left( \etav -\mathscr{G}\left(\xiv |   \tilde{\ov}  \right)\right)\left( \etav -\mathscr{G}\left(\xiv |  \tilde{\ov} \right)\right)^\TT\). 
    
    Based on \eqref{eq_lbm}, we define the position error lower bound under the mismatch model as
     \begin{equation}\label{eq_lb}
    \text{LB}\triangleq\sqrt{\text{tr}\big(\big[\text{LBM}(\xiv_0, \bar{\xiv})\big]_{1:3,1:3}\big)}.
    \end{equation}
    The \ac{ue} position \ac{rmse} is defined as $\text{RMSE} = \sqrt{\mathbb{E}\{\|\hat{\mathbf{\pv}}-\mathbf{\pv}\|_2^2\}}$, and $\text{RMSE} \geq \text{LB}$ holds.

    \subsection{Positioning Performance Under Orbital Mismatch}\label{subsec_pos_performance}
    The above \ac{mcrb} derivation provides a tractable approach to quantify the estimation error bound under the considered model mismatch due to orbital errors. We next present a numerical example to illustrate how orbit mismatch affects positioning performance. 
    To illustrate this, we compare the \ac{crb} derived under perfect orbital information with the \ac{mcrb} derived under mismatched orbital information. We specialize the previously introduced abstract model to a specific \ac{leo} downlink positioning system based on the signal model in~\cite{saleh20246g}. Specifically, we consider a scenario where the satellite is equipped with an antenna array and the \ac{ue} is equipped with a single antenna. Therefore, the observable vector consists of \acp{aod}, propagation delays, and normalized Doppler shifts extracted from the received downlink signals. Over \(L\) observation epochs, the satellite position \(\pv_{\mathsf{s},\ell}\) and velocity \(\vv_{\ell}\) at epoch \(\ell\) are calculated from the available orbital information \(\tilde{\ov}\), which are mismatched with respect to the true satellite state. Under this specific system setting, the previously defined forward model can be written explicitly as
    \begin{equation}
\etav =\mathscr{G}(\xiv|\ov)
\triangleq
[\mathscr{G}_1(\xiv|\ov)^\TT,\ldots,\mathscr{G}_L(\xiv|\ov)^\TT]^\TT,
    \end{equation}
where 
\begin{equation}
\mathscr{G}_\ell(\xiv|\ov)
\triangleq
[\theta_\ell^{\mathsf{az}},\theta_\ell^{\mathsf{el}},\tau_\ell,\nu_\ell]^\TT.
\end{equation}
The individual observable components are defined below.
\subsubsection{AoD}
 Let \(\bm{\theta}_{\ell}=[\theta_{\ell}^{\mathsf{az}},  \theta_{\ell}^{\mathsf{el}}]^{  \TT}\in\mathbb{R}^2\) denote the \ac{aod} from the \ac{leo} satellite to the \ac{ue} at epoch \(\ell\). The azimuth angle $\theta_{\ell}^{\mathsf{az}}$ and elevation angle $\theta_{\ell}^{\mathsf{el}}$ are defined as
     \begin{align}
    {\theta}_{\ell}^\mathsf{az} &= \text{atan2}\big([\Rm_{\mathsf{s},\ell}^\TT(\pv-\pv_{\mathsf{s},\ell})]_2,[\Rm_{\mathsf{s},\ell}^\TT(\pv-\pv_{\mathsf{s},\ell})]_1\big), \label{eq_thetaaz}\\
    {\theta}_{\ell}^\mathsf{el} &= \text{asin}\big([\Rm_{\mathsf{s},\ell}^\TT(\pv-\pv_{\mathsf{s},\ell})]_3/\|\pv-\pv_{\mathsf{s},\ell}\|_2\big),\label{eq_thetael}
	\end{align}
    with \(\Rm_{\mathsf{s},\ell}\) given in~\eqref{eq_rotMat}.
\subsubsection{Channel Delay}
   We define the effective delay observed at the satellite as:
    \begin{equation}\label{eq_tau_time}
      \tau_{\ell}
      = {\|\pv - \pv_{\mathsf{s},\ell}\|_2}/{c}
      + \Delta,
    \end{equation}
    where~\(c\) is the speed of light and~\(\Delta\) is the \ac{ue}-satellite clock bias. 
	\subsubsection{Doppler Shift}
    We define the normalized Doppler shifts~\(\nu_{\ell}\in\mathbb{R}\) in the \ac{ecef} frame as
    \begin{equation} \label{eq_nu}
        \nu_{\ell} = \frac{\mathbf{v}_\ell^\TT(\pv-\pv_{\mathsf{s},\ell})}{c\|\pv-\pv_{\mathsf{s},\ell}\|_2}.
	\end{equation}

Since channel delay observables are used, the \ac{ue}-satellite clock bias \(\Delta \) in~\eqref{eq_tau_time} must be estimated jointly with the \ac{ue} position. In this case, the state vector in~\eqref{eq_ue_state} is given by
\begin{equation}
    \xiv = \big[\pv^\TT,\Delta\big]^\TT.
\end{equation}

\begin{figure}
    \centering
\definecolor{mycolor1}{rgb}
{0.78039,0.13333,0.15686}%
\definecolor{mycolor2}{rgb}{0.97647,0.56078,0.20392}%
\definecolor{mycolor3}{rgb}{0.07843,0.31765,0.48627} %
\definecolor{mycolor6}{rgb}{0.41961,0.59608,0.76863}%
\definecolor{mycolor7}{rgb}{0.41569,0.68627,0.17647}%
\definecolor{mycolor8}{rgb}{0.96078,0.52549,0.49804}%
\begin{tikzpicture}

\begin{axis}[%
width=2.8in,
height=2in,
at={(0in,3in)},
scale only axis=true,
enlarge x limits=false,
enlarge y limits=false,
xmin=-22,
xmax=62,
xlabel={Transmit Power [dBm]},
xlabel style={font=\footnotesize, yshift=2pt}, 
ylabel={LB [m]},
ylabel style={font=\footnotesize, yshift=-10pt}, 
ymode=log,
ymin=140,
ymax=19000,
yminorticks=true,
tick label style={font=\footnotesize},                  
axis background/.style={fill=white},                    
axis lines=box,                                         
grid style={dotted, draw=gray!50},                      
xmajorgrids,
xminorgrids,
ymajorgrids,
yminorgrids,
legend style={
    at={(1,1)},
    anchor=north east,
    legend columns=1,
    draw=black,                                       
    font=\footnotesize,                                
    /tikz/every even column/.append style={column sep=0.5em},
    nodes={scale=0.75, transform shape}                 
},
]

\addplot [color=mycolor3, dashed, line width=1.2pt, mark size= 3pt, mark=diamond, mark options={solid, mycolor3}]
  table[row sep=crcr]{%
-20	18075.0098047851\\
-10	10231.1584540063\\
0	5871.1990204114\\
10	3500.4066894267\\
20	2289.91082727781\\
30	1738.92571927271\\
40	1522.62309326283\\
50	1440.87475935063\\
60	1420.14850842591\\
};
\addlegendentry{\text{MCRB}\((L = 6)\)}

\addplot [color=mycolor3, solid, line width=1.2pt]
  table[row sep=crcr]{%
-20	18019.9569407727\\
-10	10133.366472253\\
0	5698.41073186647\\
10	3204.45183994918\\
20	1801.99569518156\\
30	1013.33664651811\\
40	569.841073054748\\
50	320.445184183129\\
60	180.199569535967\\
};
\addlegendentry{\text{CRB}\((L = 6)\)}

\addplot [color=mycolor2, dashed, line width=1.2pt,mark size= 2.3pt , mark=square, mark options={solid, mycolor2}]
  table[row sep=crcr]{%
-20	8865.32896730703\\
-10	5119.30168443008\\
0	3106.50803370643\\
10	2104.58651500107\\
20	1666.6274863217\\
30	1499.10164910456\\
40	1435.94949092482\\
50	1426.29703660002\\
60	1419.97140970408\\
};
\addlegendentry{\text{MCRB}\((L = 10)\)}

\addplot [color=mycolor2, solid, line width=1.2pt]
  table[row sep=crcr]{%
-20	8752.17264775114\\
-10	4921.70836513364\\
0	2767.68000399932\\
10	1556.38084112252\\
20	875.217264765081\\
30	492.170836525263\\
40	276.768000401292\\
50	155.638084131094\\
60	87.5217264717647\\
};
\addlegendentry{\text{CRB}\((L = 10)\)}

\addplot [color=mycolor1, dashed, line width=1.2pt, mark size= 2.5pt, mark=o, mark options={solid, mycolor1}]
  table[row sep=crcr]{%
-20	4846.70756612029\\
-10	2964.64098256456\\
0	2035.44396900819\\
10	1636.29110220468\\
20	1485.41369060814\\
30	1438.05830202317\\
40	1418.62694981449\\
50	1416.34288807551\\
60	1409.03614034154\\
};
\addlegendentry{\text{MCRB}\((L = 14)\)}

\addplot [color=mycolor1, solid, line width=1.2pt]
  table[row sep=crcr]{%
-20	4638.77972625314\\
-10	2608.57753825974\\
0	1466.91094952874\\
10	824.904647374929\\
20	463.877972605462\\
30	260.857753819247\\
40	146.691094985179\\
50	82.4904647301455\\
60	46.38779726326\\
};
\addlegendentry{\text{CRB}\((L = 14)\)}

\addplot [color=mycolor3, line width=1.2pt, dashdotted]
  table[row sep=crcr]{%
-20	1408.25805898739\\
-10	1408.25805898739\\
0	1408.25805898739\\
10	1408.25805898739\\
20	1408.25805898739\\
30	1408.25805898739\\
40	1408.25805898739\\
50	1408.25805898739\\
60	1408.25805898739\\
};
% \addlegendentry{\text{A-MCRB }\((L = 6)\)}
\draw[->, thick]
(axis cs:-10,560) -- (axis cs:-10,1090);
\node[anchor=west, font=\footnotesize] at (axis cs:-22,370) {$\sqrt{\text{tr}\big([\text{Bias}(\xiv_0)]_{1:3,1:3}\big)}$ in~\eqref{eq_lbm}}; 

\end{axis}

\end{tikzpicture}%
    \caption{Positioning RMSE bounds versus transmit power for different numbers of observation epochs \(L\in\{6,10,14\}\). The solid curves denote the \ac{crb}, and the dashed curves denote the \ac{mcrb}.}
    \label{fig_MCRB_vs_SNR_Fix_S}
\end{figure}

    With the above observable model and state definition in place, we next present numerical results with \(L\in\{6,10,14\}\) observation epochs. As shown in Fig.~\ref{fig_MCRB_vs_SNR_Fix_S}, where the orbital information is assumed to remain outdated for \(T=24\) hours, the gap between the two bounds widens as the transmit power increases. While the \ac{crb} keeps decreasing, the \ac{mcrb} converges to a bias-dominated error bound imposed by orbit mismatch. Increasing the number of observations lowers both bounds in the noise-dominated (low-\ac{snr}) regime, but in the high-power (high-\ac{snr}) regime, the \ac{mcrb} is essentially unchanged and converges to the same limit even as more observations are added. Hence, merely increasing transmit power or accumulating observations cannot reduce the positioning \ac{rmse} under orbital errors in high-\ac{snr} regions, and orbit-error correction is necessary to ensure accurate positioning. 

\begin{figure}
    \centering
\definecolor{mycolor1}{rgb}
{0.78039,0.13333,0.15686}%
\definecolor{mycolor2}{rgb}{0.96078,0.52549,0.49804}%
\definecolor{mycolor3}{rgb}{0.97647,0.56078,0.20392}%
\definecolor{mycolor4}{rgb}{1.00000,0.73725,0.50196}%
\definecolor{mycolor5}{rgb}{0.07843,0.31765,0.48627} %
\definecolor{mycolor6}{rgb}{0.41961,0.59608,0.76863}%
\definecolor{mycolor7}{rgb}{0.41569,0.68627,0.17647}%
\definecolor{mycolor8}{rgb}{0.96078,0.52549,0.49804}%
% \definecolor{mycolor4}{rgb}{0.32353,0.19608,0.24314}%

% \definecolor{mycolor6}{rgb}{0.82353,0.19608,0.24314}%

% \definecolor{mycolor7}{rgb}{0.24314,0.22353,0.56863}%
%
\begin{tikzpicture}
\begin{axis}[%
width=2.89in,
height=1.8in,
scale only axis=true,
enlarge x limits=false,
enlarge y limits=false,
xmin=0,
xmax=22.1,
xtick={0,4,8,12,16,20,24},
xlabel={Orbital Information Update Interval [hr]},
xlabel style={font=\footnotesize, yshift=2pt}, 
ylabel={LB [m]},
ylabel style={font=\footnotesize, yshift=-5pt}, 
ymode=linear,
ymin=3,
ymax=1200,
ytick={200,400,600,800,1000,1200,1400,1600},
yminorticks=true,
tick label style={font=\footnotesize},                  
axis background/.style={fill=white},                    
axis lines=box,                                         
grid style={dotted, draw=gray!50},                      
xmajorgrids,
xminorgrids,
ymajorgrids,
yminorgrids,
legend style={
    at={(1,1.05)},                                   
    anchor=south east,
    legend columns=3,
    draw=black,                                       
    font=\footnotesize,                                
    /tikz/every even column/.append style={column sep=2.1em},
    nodes={scale=0.9, transform shape}                 
},
]

\addplot[area legend, draw=none, fill=mycolor6, fill opacity=0.16, forget plot]
table[row sep=crcr] {%
x	y\\
0.5	1.21259936055584\\
1.5	12.007269528383\\
3	17.5983625706053\\
5	70.9224162947689\\
7	75.5539463722774\\
9.5	188.835536672332\\
14	96.2852534545386\\
18	140.015061121266\\
22	118.055458226804\\
22	1072.62765486952\\
18	1244.33527437249\\
14	1080.72012480069\\
9.5	1119.39751345008\\
7	1011.93484429354\\
5	784.835643108173\\
3	492.365346082889\\
1.5	431.899645942614\\
0.5	17.6987518268563\\
}--cycle;
\addplot [color=mycolor6, line width=1.2pt, mark size=3pt, mark=diamond, mark options={solid, mycolor6}]
  table[row sep=crcr]{%
0.5	9.45567559370609\\
1.5	221.953457735499\\
3	254.981854326747\\
5	427.879029701471\\
7	543.74439533291\\
9.5	654.116525061205\\
14	588.502689127617\\
18	692.175167746878\\
22	595.341556548163\\
};
\addlegendentry{Starlink-187}

\addplot[area legend, draw=none, fill=mycolor2, fill opacity=0.16, forget plot]
table[row sep=crcr] {%
x	y\\
0.5	2.29724464100951\\
1.5	0\\
3	0\\
5	0\\
7	0\\
9.5	0\\
11.5	0\\
14	0\\
18	0\\
22	0\\
22	453.984565241431\\
18	857.505283672111\\
14	446.361996313477\\
11.5	754.936584618174\\
9.5	170.599474497284\\
7	1187.44017417735\\
5	253.321250755419\\
3	130.145834196439\\
1.5	77.903213516054\\
0.5	18.5256995590169\\
}--cycle;
\addplot [color=mycolor2, line width=1.2pt, mark size=2.3pt, mark=square, mark options={solid, mycolor2}]
  table[row sep=crcr]{%
0.5	10.4114721000132\\
1.5	36.5101133188841\\
3	53.1701277467414\\
5	89.1514157129461\\
7	198.20078216009\\
9.5	84.5475861501102\\
11.5	184.37053917345\\
14	200.928758868109\\
18	263.91230594459\\
22	222.929080416329\\
};
\addlegendentry{Kuiper}

\addplot[area legend, draw=none, fill=mycolor7, fill opacity=0.16, forget plot]
table[row sep=crcr] {%
x	y\\
0.5	2.48347308560611\\
1.5	3.91633985627827\\
3	6.39759241182777\\
5	4.8992673738806\\
7	0\\
9.5	1.01281359045238\\
11.5	15.9146881554762\\
14	7.93827762430826\\
18	30.5388258019695\\
22	43.0594481149167\\
22	430.987463995385\\
18	422.831897041334\\
14	426.930878866212\\
11.5	175.853682003835\\
9.5	219.312991696804\\
7	232.623866073517\\
5	113.272814491758\\
3	107.455171420957\\
1.5	53.2996025884042\\
0.5	18.7891218172652\\
}--cycle;
\addplot [color=mycolor7, line width=1.2pt, mark size=2.5pt, mark=triangle, mark options={solid, mycolor7}]
  table[row sep=crcr]{%
0.5	10.6362974514357\\
1.5	28.6079712223412\\
3	56.9263819163922\\
5	59.0860409328195\\
7	107.961503726206\\
9.5	110.162902643628\\
11.5	95.8841850796556\\
14	217.43457824526\\
18	226.685361421652\\
22	237.023456055151\\
};
\addlegendentry{Starlink-082}

\addplot[area legend, draw=none, fill=mycolor5, fill opacity=0.16, forget plot]
table[row sep=crcr] {%
x	y\\
0.5	3.0402939365223\\
1.5	5.49067917520801\\
3	6.4494295947226\\
5	9.53024119824868\\
7	16.5877216783936\\
9.5	16.1571326835337\\
11.5	28.987565928405\\
14	29.5879420913285\\
18	35.7076981393768\\
22	44.3365740203759\\
22	97.2610355561034\\
18	73.8440462808428\\
14	60.4177468471295\\
11.5	51.2270290956397\\
9.5	43.5663710384947\\
7	38.0767284447491\\
5	29.1240704690826\\
3	25.6187587921582\\
1.5	17.9858623809601\\
0.5	18.6383649311315\\
}--cycle;
\addplot [color=mycolor5, line width=1.2pt, mark size=2.5pt, mark=triangle, mark options={solid, rotate=180, mycolor5}]
  table[row sep=crcr]{%
0.5	10.8393294338269\\
1.5	11.738270778084\\
3	16.0340941934404\\
5	19.3271558336656\\
7	27.3322250615713\\
9.5	29.8617518610142\\
11.5	40.1072975120223\\
14	45.002844469229\\
18	54.7758722101098\\
22	70.7988047882397\\
};
\addlegendentry{ORBCOMM}

\addplot[area legend, draw=none, fill=mycolor1, fill opacity=0.16, forget plot]
table[row sep=crcr] {%
x	y\\
0.5	0.736800136233011\\
1.5	4.98885412180563\\
3	7.52339721315046\\
5	10.3224729807487\\
7	12.577801445995\\
9.5	19.5083516354869\\
11.5	22.151863651976\\
14	22.0945268313641\\
18	27.3210948878655\\
22	36.6137396198527\\
22	83.3221134351983\\
18	68.7590486410669\\
14	55.9493299082419\\
11.5	49.331660226388\\
9.5	39.1054454102641\\
7	31.2420620080328\\
5	28.2299897097997\\
3	22.7353671000966\\
1.5	16.3708129927519\\
0.5	19.7754250555495\\
}--cycle;
\addplot [color=mycolor1, line width=1.2pt, mark size=2.5pt, mark=o, mark options={solid, mycolor1}]
  table[row sep=crcr]{%
0.5	10.2561125958912\\
1.5	10.6798335572788\\
3	15.1293821566235\\
5	19.2762313452742\\
7	21.9099317270139\\
9.5	29.3068985228755\\
11.5	35.741761939182\\
14	39.021928369803\\
18	48.0400717644662\\
22	59.9679265275255\\
};
\addlegendentry{Iridium}

\end{axis}
\end{tikzpicture}%
    \caption{Average positioning error bound versus orbital information update interval for different \ac{leo} constellations. The solid curves denote the average \ac{mcrb} values over orbital error samples derived from the historical \ac{tle} dataset, and the shaded regions indicate the spread of the resulting positioning error bounds across samples.}
    \label{fig_MCRB_vs_t}
\end{figure}

    To further evaluate the influence of stale orbital information, Fig.~\ref{fig_MCRB_vs_t} presents the average \ac{mcrb}-based positioning error bound versus the \ac{tle} update interval for different constellations, with the transmit power fixed at \unit[50]{dBm}, where the averaging is performed over orbital error samples derived from historical \ac{tle} data. As the update interval increases, the orbital information becomes outdated, and the resulting mismatch directly degrades positioning performance. Notably, as shown earlier in Fig.~\ref{fig_orberr_distribution}, update intervals longer than 12 hours are not uncommon in practice. Even at such moderate stale-information levels, the resulting positioning error is already non-negligible. When the orbital information is not updated for 12 hours, the positioning error of several constellations has already grown to the order of hundreds of meters. 
    Moreover, the shaded regions become wider as the update interval increases. This indicates that the positioning error bounds obtained from different historical \ac{tle} samples become more dispersed, which is mainly caused by the increasing variability of the underlying orbital mismatch.
   
   The results reveal that the achievable positioning accuracy is governed not only by signal quality but also by the freshness of the orbital information. As the orbital information becomes stale, the resulting mismatch raises the error floor and leads to a pronounced degradation in positioning performance. To ensure accurate positioning, improving signal quality alone is insufficient; timely orbit calibration is also required.

\section{Bayesian Orbit Calibration}\label{sec_orb_calibration}
The above results highlight the importance of \ac{leo} orbit calibration. Although orbit calibration has been extensively studied in the literature, the age-dependent statistical orbital error model established in this paper provides an additional means to improve calibration accuracy. Building on this insight, this section develops a Bayesian orbit calibration framework to exploit the learned statistics of orbital errors.

When only \ac{ue}-satellite observables are available, the orbit calibration problem is fundamentally ambiguous (see Appendix~\ref{sec_app_identify} for a proof). Therefore, additional information or constraints must be introduced to make orbit calibration feasible. Typically, stable anchors such as \ac{gnss} satellites and terrestrial \acp{bs} can be employed to track \ac{leo} satellites and calibrate their orbits. In this paper, we consider an orbit calibration process performed by $M$ known anchors, using geometric observables extracted from their received signals from a target \ac{leo} satellite.  Similar to the \ac{ue} positioning, the anchor-side observables are first extracted from the received signals in $K$ fast-time epochs during which the orbit is assumed static. Let \(\hat{\etav}_{\mathsf{a},m}\) denote the observable vector extracted at the \(m\)-th anchor over the observation window. The corresponding observation model is written as
\begin{equation}
    \hat{\etav}_{\mathsf{a},m} = {\etav}_{\mathsf{a},m} + \wv_m, 
\end{equation}
where \(\etav_{\mathsf{a},m}\) denotes the true observable vector and \(\wv_m\sim\mathcal N(\mathbf 0,\Sigmam_{\mathsf{a},m})\) denotes a zero-mean Gaussian estimation error. The corresponding observable vector is further related to the orbital element vector \(\ov\) through a forward model 
\begin{equation}
\etav_{\mathsf{a},m} = \mathscr{H}_m(\ov),
\end{equation}
Accordingly, the extracted observable vector conforms
\begin{equation}
    \hat{\etav}_{\mathsf a,m}\sim \mathcal N\!\left(\mathscr{H}_m(\ov),\Sigmam_{\mathsf a,m}\right).
\end{equation}

Based on the above observation model, we next formulate the orbit calibration problem as a Bayesian estimation problem by incorporating the learned orbital error statistics as prior information. Assuming that the observations collected from different anchors are conditionally independent given \(\ov\), the joint likelihood can be written as
\begin{equation}\label{eq_joint_likli_orb}
p\!\left(\{\hat{\etav}_{\mathsf{a},m}\}_{m=1}^M | \ov \right)
=
\prod_{m=1}^{M}
\mathcal N\!\left(
\hat{\etav}_{\mathsf{a},m};
\mathscr{H}_m(\ov),
\Sigmam_{\mathsf{a},m}
\right).
\end{equation}
The orbital error statistics characterized from historical \ac{tle} data in Section~\ref{subsec_orbital_error_statistics} are used as prior information. Accordingly, we model the conditional prior distribution of \(\ov\) given an \(\tilde{\ov}\) with age $T$ as
\begin{equation}
    \ov |\tilde{\ov}
    \sim
    \mathcal N\!\left(
    \tilde{\ov}-\muv_{\ov}(T),
    \,\Sigmam_{\ov}(T)
    \right).
\end{equation}
By combining this prior with the joint likelihood in~\eqref{eq_joint_likli_orb}, the posterior distribution of \(\ov\) is given by
\begin{equation}
p\!\left(\ov | \{\hat{\etav}_{\mathsf{a},m}\}_{m=1}^M,\tilde{\ov}\right)
\propto
p\!\left(\{\hat{\etav}_{\mathsf{a},m}\}_{m=1}^M | \ov\right)
p(\ov| \tilde{\ov}).
\end{equation}
To obtain an estimate of \(\ov\), we adopt the \ac{map} estimator, which maximizes the posterior probability density as
\begin{equation}
\hat{\ov}_{\mathsf{MAP}} 
=
\arg\max_{\ov}
\;
p\!\left(\{\hat{\etav}_{\mathsf{a},m}\}_{m=1}^M | \ov\right)
p(\ov| \tilde{\ov}).
\end{equation}
By taking the negative logarithm of the posterior and omitting the terms independent of \(\ov\), the above problem can be equivalently written as{\small
\begin{align}\label{eq_map_orbit}
\hat{\ov}_{\mathsf{MAP}}
& \!=\!
\arg\min_{\ov}\ 
\sum_{m=1}^{M}
\Big(
\hat{\etav}_{\mathsf{a},m}
\!-\!
\mathscr{H}_m(\ov)
\Big)^\TT\!\!
\Sigmam_{\mathsf{a},m}^{-1}
\Big(
\hat{\etav}_{\mathsf{a},m}
\!-\!
\mathscr{H}_m(\ov)
\Big)
\notag\\
&+
\Big(
\ov - \tilde{\ov} + \muv_{\ov}(T)
\Big)^\TT
\Sigmam_{\ov}^{-1}(T)
\Big(
\ov - \tilde{\ov} + \muv_{\ov}(T)
\Big).
\end{align}}

The first term in~\eqref{eq_map_orbit} enforces consistency between the candidate orbital element vector and the observables collected from the anchors. The second term regularizes the solution according to the learned orbital error statistics associated with the stale orbit \(\tilde{\ov}\). Therefore, the proposed \ac{map} estimator jointly exploits the geometric constraints provided by the anchors and the prior statistical knowledge. Note that this formulation provides a general Bayesian framework for orbit calibration. A concrete case study is presented in the following sections to demonstrate its application and effectiveness.

\section{Case Study}\label{sec_CS}
In this section, we present a calibration-then-positioning case study in which the proposed Bayesian calibration framework is first used to calibrate the satellite orbit, followed by \ac{ue} positioning based on the calibrated orbital information. 

\begin{figure}
    \centering
    \includegraphics[width=1\linewidth]{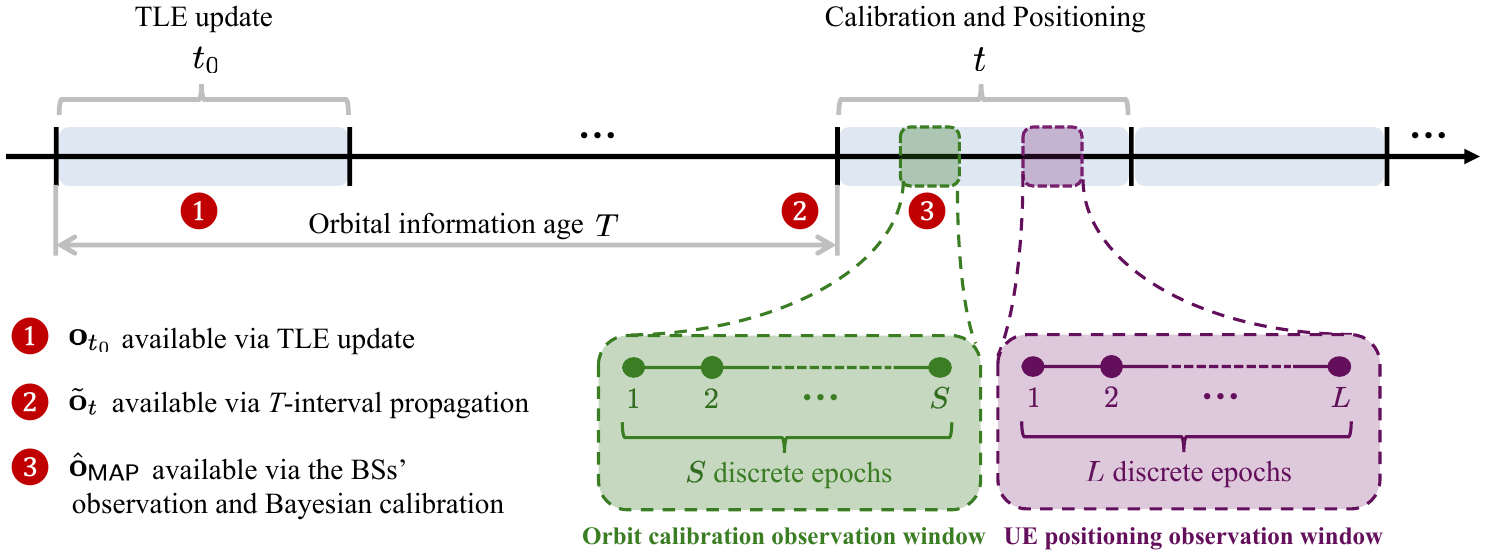}
    \caption{Timeline of the considered calibration-then-positioning scenario.}
    \label{fig_time_iterval}
\end{figure}

Specifically, we consider the scenario illustrated in Fig.~\ref{fig_time_iterval}. A \ac{tle} update is assumed to occur at the slow-time epoch \(t_0\), such that the true orbital element vector \(\ov_{t_0}\) is available. Orbit calibration and subsequent \ac{ue} positioning are both performed at a later slow-time epoch \(t\), where \(t-t_0=T\) is the orbital information age. During orbit calibration and \ac{ue} positioning, the received signals are sampled at \(K\) and \(L\) fast-time epochs, respectively, to obtain the corresponding geometric observables. Since both procedures are carried out within the same slow-time epoch \(t\), the satellite orbit is assumed to remain unchanged throughout. This setting is adopted solely for the purpose of the case study. In general, orbit calibration and \ac{ue} positioning may be separated by a non-negligible slow-time interval, during which the satellite orbit continues to evolve.

\subsection{Bayesian Orbit Calibration}\label{sub_sec_bayesian_orb}
To provide a concrete example of the general Bayesian calibration framework developed in Section~\ref{sec_orb_calibration}, we now specialize the anchor model. In this case study, the anchors are assumed to be fixed terrestrial \acp{bs} equipped with antenna arrays and with known positions and orientations. 

We focus on four types of observables, namely, \acp{aod}, \acp{aoa}, channel delays, and Doppler shifts.
For each \ac{bs} \(m\), we define its observable vector as
\begin{equation}
    \etav_{\mathsf{a},m}
\triangleq
\big[
\etav_{\mathsf{a},m,1}^\TT,\ldots,\etav_{\mathsf{a},m,K}^\TT
\big]^\TT
\in\mathbb{R}^{6K},
\end{equation}
where the \(k\)-th observable vector is given by
\begin{equation}
    \etav_{\mathsf{a},m,k}
\triangleq
\big[
\theta_{\mathsf{a},m,k}^{\mathsf{az}},
\theta_{\mathsf{a},m,k}^{\mathsf{el}},
\psi_{\mathsf{a},m,k}^{\mathsf{az}},
\psi_{\mathsf{a},m,k}^{\mathsf{el}},
\tau_{\mathsf{a},m,k},
\nu_{\mathsf{a},m,k}
\big]^\TT.
\end{equation}

Here, \(\thetav_{\mathsf{a},m,k}^{\mathsf{az}},\thetav_{\mathsf{a},m,k}^{\mathsf{el}},\tauv_{\mathsf{a},m,k}\), and \(\nuv_{\mathsf{a},m,k}\) are defined analogously to the \ac{ue} case in Section~\ref{subsec_pos_performance}, with the \ac{ue} position replaced by the \ac{bs} position \(\pv_{\mathsf{a},m}\), while \(\psiv_{\mathsf{a},m,k}^{\mathsf{az}}\) and \(\psiv_{\mathsf{a},m,k}^{\mathsf{el}}\) denote the \acp{aoa} measured at the receive array of the \(m\)-th \ac{bs}. In the local \ac{enu} frame associated with the \ac{bs} at each epoch \(k\), they are given by
{\small
\begin{align}
{\psi}_{{\mathsf{a},m,k}}^{\mathsf{az}}
&\!=\!
\text{atan2}
\Big(\!
[\Rm_{\mathsf{a},m}^{\TT}(\pv_{\mathsf{s},k}-\pv_{\mathsf{a},m})]_2,\!
[\Rm_{\mathsf{a},m}^{\TT}(\pv_{\mathsf{s},k}-\pv_{\mathsf{a},m})]_1\!
\Big), \label{eq_psiaz}\\
{\psi}_{{\mathsf{a},m,k}}^{\mathsf{el}}
&\!=\!
\text{asin}
\Big(\!
[\Rm_{\mathsf{a},m}^{\TT}(\pv_{\mathsf{s},k}-\pv_{\mathsf{a},m})]_3
/
\|\pv_{\mathsf{s},k}-\pv_{\mathsf{a},m}\|_2\!
\Big),\label{eq_psiel}
\end{align} }
where \(\Rm_{\mathsf{a},m} \) is the rotation matrix defined in~\eqref{eq_Rn2e}.
Then the extracted noisy observable vector is modeled as
\begin{equation}
\hat{\etav}_{\mathsf{a},m}
\sim
\mathcal N\!\left(
\mathscr{H}_m(\ov),
\Sigmam_{\mathsf{a},m}
\right),
\end{equation}
where
\begin{equation}
\mathscr{H}_m(\ov)
\triangleq
\big[
\etav_{\mathsf{a},m,1}^\TT,\ldots,\etav_{\mathsf{a},m,K}^\TT
\big]^\TT.
\end{equation}

With the above observation model, the general Bayesian calibration framework developed in Section~\ref{sec_orb_calibration} can be directly applied by substituting \(\mathscr{H}_m(\ov)\) into the corresponding likelihood and estimator formulations. Therefore, the \ac{map} estimator for this case study follows directly from~\eqref{eq_map_orbit}.

\subsection{UE Positioning Based on the Calibrated Orbit}\label{sub_sec_pos}
    After orbit calibration, we treat the calibrated orbital element vector \(\hat{\ov}_\mathsf{MAP}\) as known and estimate the \ac{ue} state \(\xiv\), i.e., solving~\eqref{eq_pos_only} with \(\tilde{\ov} = \hat{\ov}_\mathsf{MAP}\). 
    In addition to the \ac{mcrb} analysis, we further develop a practical \ac{ue} position estimator to more comprehensively evaluate the positioning accuracy achievable using the Bayesian-calibrated orbital element vector \(\hat{\ov}_\mathsf{MAP}\). Since the positioning problem in~\eqref{eq_pos_only} is nonlinear, we employ a two-stage estimator. First, a linear \ac{ls} problem based on the \ac{aod} and channel delay models in Section~\ref{subsec_pos_performance} is solved to obtain a coarse estimate of the \ac{ue} state. This estimate is then used to initialize a Levenberg--Marquardt algorithm for solving~\eqref{eq_pos_only}.   

   At each observation epoch~\(\ell\), the satellite-\ac{ue} propagation delay yields a pseudo-range
    \begin{equation}\label{eq_r_t}
        r_\ell \;=\; \|\pv - \pv_{\mathsf{s},\ell}\|_2 + c\Delta.
    \end{equation}
    To eliminate the unknown clock bias, we form range differences relative to a reference epoch~\(\ell=1\). For \(\ell =2,3,\dots,L\), define:
    \begin{equation}
    \epsilon_\ell = r_\ell - r_1 = d_\ell - d_1 = \|\pv - \pv_{\mathsf{s},\ell}\|_2 - \|\pv - \pv_{\mathsf{s},1}\|_2,
    \end{equation}
    where \( d_\ell = \|\pv - \pv_{\mathsf{s},\ell}\|_2\) denotes the true geometric distance.
    This range difference defines a \ac{3d} hyperboloid with foci at \(\pv_{\mathsf{s},1}\) and \(\pv_{\mathsf{s},\ell}\), where the \ac{ue} lies at the intersection of \(L-1\) such surfaces. However, directly intersecting multiple hyperboloids is computationally challenging. Instead, we reformulate it into a linear system by subtracting squared ranges:
        \begin{equation}
    \|\pv - \pv_{\mathsf{s},\ell}\|_2^2 -\|\pv - \pv_{\mathsf{s},1}\|_2^2  \;=\; (d_1 + \epsilon_\ell)^2 -d_1^2.
    \end{equation}
    After expanding and rearranging, the linearized equations are denoted as
    \begin{equation} \label{eq_inital_LS}
    \pv^\TT(\pv_{\mathsf{s},\ell} - \pv_{\mathsf{s},1})
        + d_1  \epsilon_\ell
        = \frac{1}{2}\Bigl(\|\pv_{\mathsf{s},\ell}\|_2^2 - \|\pv_{\mathsf{s},1}\|_2^2 - \epsilon_\ell^2\Bigr).
    \end{equation}
    Stacking these equations for epochs \(\ell=2,\dots,L\) in matrix form, we can define\footnote{ Actually, \(d_1 = \|\pv - \pv_{\mathsf{s},1}\|_2\), so~\(d_1\) is not an independent variable but a function of~\(\pv\). However, since this is a coarse estimation, we relax~\(\pv,d_1\) as two independent unknowns in order to cast the problem into a linear problem.}
    {\small \begin{equation}
        \Hm_\mathsf{d}  \!\!  =   \! \! \!  
    \begin{bmatrix}\! \! 
       (\pv_{\mathsf{s},2}  \! -\! \!   \pv_{\mathsf{s},1})^\TT,\! \! \!  &          \epsilon_2  \\
       \! \! (\pv_{\mathsf{s},3} \!  - \! \!  \pv_{\mathsf{s},1})^\TT,\! \! \!  &        \epsilon_3  \\
      \vdots &        \vdots \\
       \! (\pv_{\mathsf{s},L}  \! - \! \!  \pv_{\mathsf{s},1})^\TT, \! \! \! &        \epsilon_L
    \end{bmatrix}  , 
      \cv_\mathsf{d}   =   
    \frac{1}{2}   
    \begin{bmatrix}
       \|\pv_{\mathsf{s},2}\|_2^2  -  \|\pv_{\mathsf{s},1}\|_2^2  -  \epsilon_2^2  \\
       \|\pv_{\mathsf{s},3}\|_2^2  -  \|\pv_{\mathsf{s},1}\|_2^2  -  \epsilon_3^2  \\
      \vdots \\
       \|\pv_{\mathsf{s},L}\|_2^2  -  \|\pv_{\mathsf{s},1}\|_2^2 - \epsilon_L^2 
    \end{bmatrix},  \notag
    \end{equation} }
    \begin{equation}
                \qv   =  \begin{bmatrix}
        \pv^\TT , d_1
        \end{bmatrix}^\TT.
    \end{equation}
    Hence, we have $\Hm_\mathsf{d}\qv=\cv_\mathsf{d}$.

    In addition to the time delay, the \acp{aod} provide directional information on the \ac{ue} location. The AoD pair \([\theta_\ell^{\mathsf{az}},\theta_\ell^{\mathsf{el}}]\), obtained in the \ac{sl} frame, is mapped to a unit direction vector in \ac{ecef} frame as \( \sv_\ell \!\!=\!\! \Rm_{\mathsf{s},\ell}[ \cos \theta^\mathsf{el}_\ell \cos \theta^\mathsf{az}_\ell, \cos \theta^\mathsf{el}_\ell \sin \theta^\mathsf{az}_\ell,\sin \theta^\mathsf{el}_\ell ]^\TT\). The orthogonal projector onto the plane normal to \(\sv_\ell\) is \(\Pm_\ell = \Id_3 - \sv_\ell \sv_\ell^\TT \in\mathbb{R}^{3\times3}.\) Geometrically, \(\Pm_\ell(\pv - \pv_{\mathsf{s},\ell})=0\) forces \(\pv\) onto the \ac{los} ray from \(\pv_{\mathsf{s},\ell}\). We stack these relations over epochs into the AoD \ac{ls} system, given by
\begin{equation}
    \Hm_\theta=
\begin{bmatrix}
\Pm_1 ,& \mathbf{0}\\
\Pm_2 ,& \mathbf{0}\\
\vdots\\
\Pm_L ,& \mathbf{0}
\end{bmatrix}
,
\qquad
\cv_\theta=
\begin{bmatrix}
\Pm_1\pv_{\mathsf{s},1} \\
\Pm_2\pv_{\mathsf{s},2}\\
\vdots\\
\Pm_L\pv_{\mathsf{s},L}
\end{bmatrix}.
\end{equation}
    This leads to the AoD-based linear system \(\Hm_\theta \qv = \cv_\theta\).
    
    We finally combine the two sets of observables into a unified weighted \ac{ls} problem as:
    \begin{equation}
    \hat{\qv} = \arg \min_{\qv}\ \bigl(\Hm\qv - \cv\bigr)^\TT \Wm \bigl(\Hm\qv - \cv\bigr),
    \end{equation}
    where \(\Hm = [\Hm_\mathsf{d}^\TT,\Hm_\theta^\TT]^\TT\) and \(\cv = [\cv_\mathsf{d}^\TT,\cv_\theta^\TT]^\TT\) and the weighting matrix \(\Wm\) is chosen diagonal, assigning unit weight to delay rows and inverse-range weight to AoD rows as \(\Wm=\mathrm{diag} \big(\mathbf{1}_{L-1}^\TT,\frac{1}{r_1}\mathbf{1}_{3L}^\TT)\in \mathbb{R}^{(4L-1)\times (4L-1)}\).
    The closed-form solution is given by:
    \begin{equation} \label{eq_init_close_sol}
    \hat \qv = (\Hm^\TT \Wm \Hm)^{-1}\Hm^\TT \Wm \cv.
    \end{equation}
Finally, we extract the coarse estimates \(\hat{\pv}_0=[\hat{\qv}]_{1:3}\) and \(\hat{\Delta}_0=r_1-[\hat{\qv}]_4\) for the \ac{ue} position and clock bias, respectively. These estimates are then used to initialize the subsequent nonlinear \ac{ue} positioning based on the calibrated orbit.

    With the coarse estimates \((\hat{\pv}_{0},\hat{\Delta}_{0})\) available, we then solve the following weighted nonlinear least-squares problem formulated in~\eqref{eq_pos_only} for \ac{ue} positioning. The optimization is carried out using a Levenberg-Marquardt method with analytic Jacobians. The whole estimator is summarized in Algorithm~\ref{alg_pos}.
   \begin{algorithm}[t]
\caption{Bayesian Orbit Calibration and UE Positioning}
\label{alg_pos}
\begin{algorithmic}[1]
\Statex \textbf{Inputs:} 
Anchor observables \(\{\hat{\etav}_{\mathsf{a},m}\}_{m=1}^M\), \ac{ue} observables \(\hat{\etav}\), propagation interval \(T\), orbital mismatch statistics \((\muv_{\ov}(T),\Sigmam_{\ov}(T))\), and prior orbital elements \(\ov_{t_0}\).
\Statex \textbf{Outputs:} Calibrated orbit \(\hat{\ov}_{\mathsf{MAP}}\) and \ac{ue} position \(\hat{\pv}\).

\State \textbf{Orbit propagation:} Predict the orbital elements at time \(t \) as \(\tilde{\mathbf{o}}_t=\mathscr{F}(\mathbf{o}_{t_0},T)\).

\State \textbf{Bayesian orbit calibration:} Estimate \(\hat{\ov}_{\mathsf{MAP}}\) from the anchor observations \(\{\hat{\etav}_{\mathsf{a},m}\}_{m=1}^M\) via~\eqref{eq_map_orbit} given \((\muv_{\ov}(T),\Sigmam_{\ov}(T))\) and \(\tilde{\ov}_{t}\).

\State \textbf{Coarse \ac{ue} state estimation:} Obtain a coarse estimate \(\hat{\pv}_0\) via~\eqref{eq_r_t}--\eqref{eq_init_close_sol} using the propagation delays and AoDs extracted from \(\hat{\etav}\).

\State \textbf{\ac{ue} positioning under the calibrated orbit:} Solve~\eqref{eq_pos_only} using \(\hat{\ov}_{\mathsf{MAP}}\) and the initialization \(\hat{\pv}_0\).
\end{algorithmic}
\end{algorithm}

   \subsection{Complexity Analysis}
 In the orbit-calibration stage, each of the \(M\) anchors provides an observable vector of dimension \(KS\), where \(S\) is the number of observables used in each epoch. Hence, the total number of observations is \(MKS\). The \ac{map} estimator proposed in Section~\ref{sec_orb_calibration} is solved iteratively using a gradient-based optimization method. In each iteration, residual evaluation requires \(\mathcal{O}(MKS)\) operations. The Jacobian matrix has size \((MKS)\)-by-\(6\), and constructing the analytical Hessian requires \(\mathcal{O}(MKS)\) operations. Solving the resulting \(6\)-by-\(6\) system incurs only a constant cost. Therefore, the per-iteration complexity is \(\mathcal{O}(MKS)\).

\section{Simulation Results}\label{sec_simumation}

This section presents simulation results for the case studies in Section~\ref{sec_CS}. Throughout the simulations, we use a representative satellite from the Starlink-082 constellation, with the corresponding orbital elements \([a,e,i,\Omega,\omega,\gamma]=[6945\ \mathrm{km},\,0.0003,\,70^\circ,\,223^\circ,\,262^\circ,\,98^\circ]\) collected on 17/03/2026. The mean vector and covariance matrix of orbital errors with \(T = 24\) are used and presented in~\eqref{eq_mu_o_app} and~\eqref{eq_sigma_o_app}, respectively. We adopt the \ac{ofdm} signal model in~\cite{saleh20246g}, and the parameters in Table~\ref{table_system_params} are used as the simulation settings for positioning-observable extraction.

\begin{figure*}[!t]
\centering
{\footnotesize
\begin{minipage}[t]{0.2\textwidth}
\begin{equation}
\muv_{\mathsf{o}}=
\begin{bmatrix}
-5.9\times 10^{-3}\\
-7.5\times 10^{-6}\\
-2.0\times 10^{-7}\\
-2.8\times 10^{-5}\\
\;\;9.7\times 10^{-3}\\
-9.4\times 10^{-3}
\end{bmatrix}
\label{eq_mu_o_app}
\end{equation}
\end{minipage}
\hfill
\begin{minipage}[t]{0.75\textwidth}
\begin{equation}
\Sigmam_{\mathsf{o}}=
\begin{bmatrix}
 3.3\times 10^{-3} &  2.7\times 10^{-7} & -7.0\times 10^{-8} &  8.0\times 10^{-8} & -2.5\times 10^{-4} &  1.7\times 10^{-4} \\
 2.7\times 10^{-7} &  1.8\times 10^{-9} &  1.2\times 10^{-10} &  4.5\times 10^{-10} & -6.6\times 10^{-7} &  6.5\times 10^{-7} \\
-7.0\times 10^{-8} &  1.2\times 10^{-10} &  3.1\times 10^{-10} &  1.3\times 10^{-10} & -5.4\times 10^{-8} &  5.5\times 10^{-8} \\
 8.0\times 10^{-8} &  4.5\times 10^{-10} &  1.3\times 10^{-10} &  1.3\times 10^{-9} & -4.6\times 10^{-7} &  4.6\times 10^{-7} \\
-2.5\times 10^{-4} & -6.6\times 10^{-7} & -5.4\times 10^{-8} & -4.6\times 10^{-7} &  1.3\times 10^{-3} & -1.2\times 10^{-3} \\
 1.7\times 10^{-4} &  6.5\times 10^{-7} &  5.5\times 10^{-8} &  4.6\times 10^{-7} & -1.2\times 10^{-3} &  1.2\times 10^{-3}
\end{bmatrix}
\label{eq_sigma_o_app}
\end{equation}
\end{minipage}
}
\end{figure*}

\begin{table}[t] 
    \renewcommand{\arraystretch}{1.2}
    \caption{Simulation Parameters}
    \vspace{-1em}
    \label{table_system_params}
    \centering
    \begin{tabular}{ll}
    \hthickline
    \textbf{Parameter} &\textbf{Value} \\
    \hline
    Center frequency      & \unit[12.7]{GHz} \\
    Subcarrier bandwidth   & \unit[120]{kHz} \\
    Speed of light         & \(3\times 10^8\) m/s \\
    \ac{cp} duration      & \unit[0.58]{us} \\
    Number of subcarriers        & 1024 \\
    Number of transmissions     & 256 \\
    Noise \ac{psd}& \unit[-174]{dBm/Hz} \\

    \hthickline
    \end{tabular}
    \vspace{-1em}
    \end{table} 

\subsection{Impact Evaluation of Orbital Errors}
    In this subsection, we quantify how \ac{leo} orbital errors propagate into the observables~\(\etav\) and assess their sensitivity with respect to orbital element errors. 

\subsubsection{Observable Sensitivity Analysis}

    \begin{table*}[t]
          \centering
          \caption{Sensitivity of Observables to Orbital Elements}

          \setlength{\tabcolsep}{4.5pt} 
          \renewcommand{\arraystretch}{1.2} 
          \label{tab_sensitivity}
          \begin{tabular}{c c c c c c c c}
            \hline
            Observable & Semi-major Axis $(a)$  & Eccentricity $(e)$ & Inclination $(i)$ & RAAN $(\Omega)$ & Argument of Periapsis $(\omega)$ &True anomaly  $(\gamma)$ & $\omega+ \gamma$ \\
            \hline 
             $\theta^{\mathsf{az}}$   
              & $1.6\times10^{-7}$ 
              & $5.8\times10^{-10}$ 
              & $3.8\times10^{-8}$ 
              & $1.2\times10^{-6}$ 
              & $1.2\times10^{-3}$ 
              & $1.6\times10^{-4}$
              & $3.1\times10^{-5}$\\
            $\theta^{\mathsf{el}}$   
              & $9.2\times10^{-7}$
              & $4.9\times10^{-6}$ 
              & $4.9\times10^{-8}$ 
              & $1.4\times10^{-6}$ 
              & $1.3\times10^{-3}$ 
              & $1.3\times10^{-4}$
              & $3.2\times10^{-5}$\\
            $\tau$  
              & $1.6\times10^{-5}$ 
              & $6.8\times10^{-5}$ 
              & $2.6\times10^{-6}$ 
              & $7.5\times10^{-5}$       
              & $7.2\times10^{-2}$    
              & $4.7\times10^{-3}$
              & $1.8\times10^{-3}$\\  
            $\nu$    
              & $2.5\times10^{-5}$ 
              & $6.3\times10^{-5}$ 
              & $5.2\times10^{-6}$   
              & $1.5\times10^{-4}$     
              & $1.6\times10^{-1}$      
              & $2.1\times10^{-2}$
              & $4.1\times10^{-3}$\\
            \hline
          \end{tabular}
          \vspace{-1em}
        \end{table*}
 
To further evaluate how orbital element errors propagate to the observables, we perform a covariance-weighted sensitivity analysis based on the Jacobian of the observable vector with respect to the orbital element vector. Specifically, we define
\begin{equation}
    \Sm \triangleq \Dm_{\etav}^{-1}\Sm_0\Sigmam_{\ov},
\end{equation}
where
\begin{equation}
   \Sm_0 \triangleq \frac{\partial \etav}{\partial \ov}\in\mathbb{R}^{4L\times 6}.
\end{equation}
Here, \(\Dm_{\etav}\in\mathbb{R}^{4L\times 4L}\) is the observable normalization matrix defined as
\begin{equation}
\Dm_{\etav}= \mathrm{diag}(\etav_{\mathsf{rms}})\otimes \Id_L,
\end{equation}
with
\begin{equation}
\etav_{\mathsf{rms}}
=
\frac{1}{\sqrt{L}}
\big[
\|\thetav^{\mathsf{az}}\|_2,
\|\thetav^{\mathsf{el}}\|_2,
\|\tauv\|_2,
\|\nuv\|_2
\big]^\TT
\in \mathbb{R}^{4}.
\end{equation}

Therefore, \(\Sm\) characterizes how the relative variations of the observables are influenced by orbital errors. To summarize the sensitivity behavior over the observation window, we further average the absolute value of each sensitivity entry over the \(L\) epochs. Denoting the \((i,j)\)-th entry of \(\Sm\) at epoch \(\ell\) by \(s_{ij}(\ell)\), the reported time-averaged sensitivity is
\begin{equation}
\bar{s}_{ij}
\triangleq
\frac{1}{L}\sum_{\ell=1}^{L}\left|s_{ij}(\ell)\right|.
\end{equation}
The resulting \(4\times 6\) matrix \([\bar{s}_{ij}]\) is reported in Table~\ref{tab_sensitivity}.

The results show that angular observables \(\theta^{\mathsf{az}}\) and \(\theta^{\mathsf{el}}\) are less sensitive to \(a\), \(e\), and \(i\), while being strongly affected by the angular elements \(\omega\) and \(\gamma\). This indicates that the degradation of the observables is mainly driven by phase-related angular errors, rather than by errors in the orbital size or shape. Note that for near-circular orbits, \(\omega\) and \(\gamma\) become strongly coupled, and the more meaningful quantity is the combination \(\omega+\gamma\)~\cite{vallado2001fundamentals}. As shown in the last column of Table~\ref{tab_sensitivity}, the combined variable \(\omega+\gamma\) remains the dominant factor for all observables. From a geometric perspective, \(\omega\) and \(\gamma\) jointly determine the instantaneous satellite location along the orbit, and therefore affect both the satellite-\ac{ue} distance and the line-of-sight range rate. Consequently, perturbations in these two angular elements are directly translated into delay and Doppler errors, which explains their dominant influence on the observables.

\subsection{Bayesian Orbit Calibration}
\begin{figure}
    \centering
\definecolor{mycolor1}{rgb}
{0.78039,0.13333,0.15686}%
\definecolor{mycolor2}{rgb}{0.96078,0.52549,0.49804}%
\definecolor{mycolor3}{rgb}{0.97647,0.56078,0.20392}%
\definecolor{mycolor4}{rgb}{1.00000,0.73725,0.50196}%
\definecolor{mycolor5}{rgb}{0.07843,0.31765,0.48627} %
\definecolor{mycolor6}{rgb}{0.41961,0.59608,0.76863}%
\definecolor{mycolor7}{rgb}{0.41569,0.68627,0.17647}%
\definecolor{mycolor8}{rgb}{0.96078,0.52549,0.49804}%

\begin{tikzpicture}
\begin{axis}[%
width=2.8in,
height=2.2in,
scale only axis=true,
enlarge x limits=false,
enlarge y limits=false,
xmin=0.95,
xmax=4.1,
ymin=0,
ymax=0.14,
xtick={1,2,3,4},
xticklabels={1,2,3,4},
xlabel={Number of Calibration BSs $M$},
xlabel style={font=\footnotesize, yshift=2pt}, 
ylabel={Positioning RMSE [m]},
ylabel style={font=\footnotesize, yshift=-6pt},
ymode=log,
scaled y ticks=false,
ytick={0.001,0.002,0.004,0.008,0.016,0.032,0.064,0.128},
yticklabels={0.01,0.02,0.04,0.08,0.16,0.32,0.64,1.28},
tick label style={font=\footnotesize},
extra description/.code={
    \node[anchor=south west,font=\footnotesize]
    at (axis description cs:0,1.02)
    {$\times 10^{-1}$};
},
ylabel={Semi-Major-Axis Error $|\hat a-a_{\mathsf{true}}|$ [km]},
axis background/.style={fill=white},                    
axis lines=box,                                         
grid style={dotted, draw=gray!50},                      
xmajorgrids,
xminorgrids,
ymajorgrids,
yminorgrids,
legend style={
at={(1,1)},
    anchor=north east,
    legend columns=1,
    draw=black,                                       
    font=\footnotesize,                                
    /tikz/every even column/.append style={column sep=0.5em},
    nodes={scale=0.85, transform shape}                 
},
]

 \addplot [color=mycolor7, dashed, line width=1pt, mark size=2pt, mark=triangle, mark options={solid, mycolor7}]
  table[row sep=crcr]{%
1	0.108783458374802\\
2	0.0816183015814147\\
3	0.00209138814386733\\
4	0.00184185954472582\\
};
 \addlegendentry{ML ($T$ = 24 hr)}

% \addplot [color=mycolor3, dashed, line width=1.0pt, mark size=2pt, mark=diamond, mark options={solid, mycolor3}]
%   table[row sep=crcr]{%
% 1	0.098460387139113\\
% 2	0.0724997347849978\\
% 3	0.00232932933539587\\
% 4	0.00207136960054299\\};
% \addlegendentry{ML ($T$ = 12 hr)}

\addplot [color=mycolor1, dashed, line width=1.0pt, mark size=2pt, mark=square, mark options={solid, mycolor1}]
  table[row sep=crcr]{%
1	0.0810678390019461\\
2	0.0582624658177053\\
3	0.00186909578641803\\
4	0.0016566980111179\\};
\addlegendentry{ML ($T$ = 5 hr)}

 \addplot [color=mycolor5, dashed, line width=1pt, mark size=2.0pt, mark=o, mark options={solid, mycolor5}]
  table[row sep=crcr]{%
1	0.0519928466569297\\
2	0.0387294130086104\\
3	0.00142665344529291\\
4	0.00133861662959589\\ };
\addlegendentry{ML ($T$ = 1 hr)}

\addplot [color=mycolor7, solid, line width=1pt, mark size=2pt, mark=triangle, mark options={solid, mycolor7}]
  table[row sep=crcr]{%
1	0.101746120333047\\
2	0.0175720273492743\\
3	0.0020682621083381\\
4	0.00183254015297507\\ };
\addlegendentry{MAP ($T$ = 24 hr)}

%  \addplot [color=mycolor3, solid, line width=1.0pt, mark size=2pt, mark=diamond, mark options={solid, mycolor3}]
%   table[row sep=crcr]{%
% 1	0.0964522944218231\\
% 2	0.0178123755346746\\
% 3	0.00231598965169155\\
% 4	0.00206680980352303\\ };
% \addlegendentry{MAP ($T$ = 12 hr)}

 \addplot [color=mycolor1, solid, line width=1.0pt, mark size=2pt, mark=square, mark options={solid, mycolor1}]
  table[row sep=crcr]{%
1	0.057012799280642\\
2	0.0143321568462227\\
3	0.00188202094389567\\
4	0.001638507626609\\};
 \addlegendentry{MAP ($T$ = 5 hr)}

\addplot [color=mycolor5, solid, line width=1pt, mark size=2.0pt, mark=o, mark options={solid, mycolor5}]
  table[row sep=crcr]{%
1	0.0154997341111311\\
2	0.00703308413162063\\
3	0.0013943565349166\\
4	0.000971029507581989\\};
 \addlegendentry{MAP ($T$ = 1 hr)}

\end{axis}

\end{tikzpicture}%
    \caption{Semi-major-axis estimation error versus the number of BSs used for orbit calibration under different orbital information ages. Dashed and solid lines denote the ML and MAP estimates, respectively. }
    \label{fig_orb_pram_corrected}
\end{figure}
Based on the proposed ground-\ac{bs} observation model in Section~\ref{sub_sec_bayesian_orb}, we evaluate the proposed orbit calibration framework numerically. Specifically, we compare the calibration performance of the proposed \ac{map} and conventional \ac{ml} estimators as a function of the number of \acp{bs} under different orbital information age \(T\). As a representative example, we focus on the estimation of the semi-major axis \(a\). As shown in Fig.~\ref{fig_orb_pram_corrected}, increasing the number of \acp{bs} used for orbit calibration generally improves the estimation accuracy. The proposed \ac{map} estimator consistently outperforms the \ac{ml} benchmark. This advantage is especially pronounced when only one or two \acp{bs} are used. 
As the number of \acp{bs} increases, the performance gap between \ac{map} and \ac{ml} gradually narrows. The result indicates that the learned orbital error statistics are particularly beneficial in anchor-limited scenarios.

\subsection{UE Positioning}

    \begin{figure}
    \centering
\definecolor{mycolor1}{rgb}
{0.78039,0.13333,0.15686}%
\definecolor{mycolor2}{rgb}{0.96078,0.52549,0.49804}%
\definecolor{mycolor3}{rgb}{0.97647,0.56078,0.20392}%
\definecolor{mycolor4}{rgb}{1.00000,0.73725,0.50196}%
\definecolor{mycolor5}{rgb}{0.07843,0.31765,0.48627} %
\definecolor{mycolor6}{rgb}{0.41961,0.59608,0.76863}%
\definecolor{mycolor7}{rgb}{0.41569,0.68627,0.17647}%
\definecolor{mycolor8}{rgb}{0.96078,0.52549,0.49804}%
% \definecolor{mycolor4}{rgb}{0.32353,0.19608,0.24314}%

% \definecolor{mycolor6}{rgb}{0.82353,0.19608,0.24314}%

% \definecolor{mycolor7}{rgb}{0.24314,0.22353,0.56863}%
%
\begin{tikzpicture}
\begin{axis}[%
width=2.8in,
height=2.2in,
scale only axis=true,
enlarge x limits=false,
enlarge y limits=false,
xmin=0.95,
xmax=4.1,
xtick={1,2,3,4},
xticklabels={1,2,3,4},
xlabel={Number of Calibartion BSs $M$},
xlabel style={font=\footnotesize, yshift=2pt}, 
ylabel={Positioning RMSE [m]},
ylabel style={font=\footnotesize, yshift=-10pt}, 
ymode=log,
ymin=30,
ymax=500,
yminorticks=true,
tick label style={font=\footnotesize},                  
axis background/.style={fill=white},                    
axis lines=box,                                         
grid style={dotted, draw=gray!50},    
ymode=log,
xmajorgrids,
xminorgrids,
ymajorgrids,
yminorgrids,
ytick={30,50,100,200,500},
yticklabels={30,50,100,200,500},
legend style={
at={(0.97,0.7)},
    anchor=north east,
    legend columns=1,
    draw=black,                                   
    font=\footnotesize,                                
    /tikz/every even column/.append style={column sep=0.5em},
    nodes={scale=0.85, transform shape}                 
},
]
% \addplot [color=mycolor7, dashed, line width=1pt, mark size=2pt, mark=triangle, mark options={solid, mycolor7}]
%  plot [error bars/.cd, y dir=both, y explicit, error bar style={line width=1pt}, error mark options={line width=1.0pt, mark size=1.0pt, rotate=90}]
%  table[row sep=crcr, y error plus index=2, y error minus index=3]{%
% 1	201.107180454077	26.8999168034977	26.8999168034977\\
% 2	162.418239191068	22.1067408484926	22.1067408484926\\
% 3	34.3236879458645	0.474818384709598	0.474818384709598\\
% 4	34.5316685062098	0.517155915216903	0.517155915216903\\
% };
% \addlegendentry{ML ($T$ = 24 hr)}

% \addplot [color=mycolor3, dashed, line width=1.0pt, mark size=2pt, mark=diamond, mark options={solid, mycolor3}]
%  plot [error bars/.cd, y dir=both, y explicit, error bar style={line width=1pt}, error mark options={line width=1pt, mark size=2pt, rotate=90}]
%  table[row sep=crcr, y error plus index=2, y error minus index=3]{%
% 1	180.841947241714	23.173171219407	23.173171219407\\
% 2	143.448156983108	18.5970470980798	18.5970470980798\\
% 3	35.2871438916221	0.496626801772454	0.496626801772454\\
% 4	35.557409670529	0.553566486154195	0.553566486154195\\
% };
% \addlegendentry{ML ($T$ = 12 hr)}

\addplot [color=mycolor1, dashdotted, line width=1pt, mark size=2.0pt, mark=square, mark options={solid, mycolor1}]
 plot [error bars/.cd, y dir=both, y explicit, error bar style={line width=1pt}, error mark options={line width=1pt, mark size=1.0pt, rotate=90}]
  table[row sep=crcr]{%
1	445.191647288501\\
2	445.191647288501\\
3	445.191647288501\\
4	445.191647288501\\
};
\addlegendentry{Uncalibrated ($T$ = 5 hr)}

\addplot [color=mycolor5, dashdotted, line width=1pt, mark size=2.0pt, mark=o, mark options={solid, mycolor5}]
 plot [error bars/.cd, y dir=both, y explicit, error bar style={line width=1pt}, error mark options={line width=1pt, mark size=1.0pt, rotate=90}]
  table[row sep=crcr]{%
1	242.19892473458\\
2	242.19892473458\\
3	242.19892473458\\
4	242.19892473458\\
};
\addlegendentry{Uncalibrated ($T$ = 1 hr)}

\addplot [color=mycolor1, dashed, line width=1.0pt, mark size=2pt, mark=square, mark options={solid, mycolor1}]
  table[row sep=crcr]{%
1	155.167575467592\\
2	134.719131418968\\
3	34.3132137442709\\
4	34.6559352480122\\
};
\addlegendentry{ML ($T$ =  5 hr)}

\addplot [color=mycolor5, dashed, line width=1pt, mark size=2.0pt, mark=o, mark options={solid, mycolor5}]
  table[row sep=crcr]{%
1	101.2676217288\\
2	87.5427539501901\\
3	32.9021021960109\\
4	33.0737460049633\\
};
\addlegendentry{ML ($T$ = 1 hr)}

% \addplot [color=mycolor7, solid, line width=1pt, mark size=2pt, mark=triangle, mark options={solid, mycolor7}]
%  plot [error bars/.cd, y dir=both, y explicit, error bar style={line width=1pt}, error mark options={line width=1.0pt, mark size=1.0pt, rotate=90}]
%  table[row sep=crcr, y error plus index=2, y error minus index=3]{%
% 1	181.013937884718	23.7359961441154	23.7359961441154\\
% 2	44.0839030711148	4.57255149631697	4.57255149631697\\
% 3	34.4527083514251	0.466632267045595	0.466632267045595\\
% 4	34.4972452593743	0.513435820825472	0.513435820825472\\
% };
% \addlegendentry{MAP ($T$ = 24 hr)}

% \addplot [color=mycolor3, solid, line width=1.0pt, mark size=2pt, mark=diamond, mark options={solid, mycolor3}]
%  plot [error bars/.cd, y dir=both, y explicit, error bar style={line width=1pt}, error mark options={line width=1pt, mark size=2pt, rotate=90}]
%  table[row sep=crcr, y error plus index=2, y error minus index=3]{%
% 1	170.802248650381	21.9710016141416	21.9710016141416\\
% 2	47.741694257972	5.36330684291807	5.36330684291807\\
% 3	35.357603466882	0.514613247980282	0.514613247980282\\
% 4	35.5375962412424	0.55837320020631	0.55837320020631\\
% };
% \addlegendentry{MAP ($T$ = 12 hr)}

\addplot [color=mycolor1, solid, line width=1.0pt, mark size=2pt, mark=square, mark options={solid, mycolor1}]
  table[row sep=crcr]{%
1	90.9343152999682\\
2	48.7836814944046\\
3	34.3624943997069\\
4	34.6903294428119\\
};
\addlegendentry{MAP ($T$ = 5 hr)}

\addplot [color=mycolor5, solid, line width=1pt, mark size=2.0pt, mark=o, mark options={solid, mycolor5}]
 plot [error bars/.cd, y dir=both, y explicit, error bar style={line width=1pt}, error mark options={line width=1pt, mark size=1.0pt, rotate=90}]
  table[row sep=crcr]{%
1	41.0748766322997\\
2	33.1883294002622\\
3	33.2600391811016\\
4	32.95327736366\\
};
\addlegendentry{MAP ($T$ = 1 hr)}

\end{axis}

\end{tikzpicture}%
    \caption{Positioning RMSE versus the number of BSs used for orbit calibration under different orbital information update intervals. Dashed and solid lines denote the ML and MAP estimators, respectively, while the dash-dotted lines denote the positioning RMSE obtained by directly using the propagated orbit without calibration.}
    \label{fig_MAP_vs_ML}
\end{figure}

    Following the orbit calibration results above, we now present the corresponding \ac{ue} positioning results using the positioning method proposed in Section~\ref{sub_sec_pos}. As shown in Fig.~\ref{fig_MAP_vs_ML}, the propagated orbit without calibration leads to the largest positioning error, which confirms the necessity of orbit calibration under orbit errors. After calibration, both the \ac{ml} and the proposed \ac{map} estimators substantially improve the final positioning accuracy. However, the proposed \ac{map} estimator consistently outperforms the \ac{ml} benchmark when the number of \acp{bs} is limited. These results show that the improved orbit estimation achieved by the proposed Bayesian calibration framework can be effectively translated into \ac{ue} positioning accuracy gains.

\begin{figure}
    \centering
\definecolor{mycolor1}{rgb}{0.78039,0.13333,0.15686}%
\definecolor{mycolor2}{rgb}{0.97647,0.56078,0.20392}%
\definecolor{mycolor3}{rgb}{0.18431,0.49804,0.75686}%
\definecolor{mycolor4}{rgb}{0.82353,0.19608,0.24314}%
\definecolor{mycolor5}{rgb}{0.3804,0.5647,0.3255}%

\begin{tikzpicture}
\begin{axis}[%
width=2.8in,
height=2.2in,
scale only axis=true,
enlarge x limits=false,
enlarge y limits=false,
xmin=-20,
xmax=82,
xlabel={Transmit Power [dBm]},
xlabel style={font=\footnotesize, yshift=2pt}, 
ylabel={Positioning RMSE [m]},
ylabel style={font=\footnotesize, yshift=-10pt}, 
ymode=log,
ymin=20,
ymax=3000000,
yminorticks=true,
tick label style={font=\footnotesize}, axis background/.style={fill=white},               
axis lines=box,                                   
grid style={dotted, draw=gray!50},                      
xmajorgrids,
xminorgrids,
ymajorgrids,
yminorgrids,
legend style={
    at={(1,1)},                               
    anchor=north east,
    legend columns=1,
    draw=black,                                    
    font=\footnotesize,                            /tikz/every even column/.append style={column sep=0.5em},
    nodes={scale=0.72, transform shape}           
},
]

\addplot [color=mycolor1, dashed, line width=1.2pt, mark size=3pt, mark=star, mark options={solid, mycolor1}]
  table[row sep=crcr]{%
-20	1078420.62245234\\
-5.71428571428571	159562.407496323\\
8.57142857142857	48505.9549962362\\
22.8571428571429	19165.2905061812\\
37.1428571428571	15604.3998685585\\
51.4285714285714	14055.0114376653\\
65.7142857142857	12978.3004487712\\
80	12919.4209835866\\
};
\addlegendentry{Linear LS initialization}

\addplot [color=mycolor2, dashed, line width=1.2pt, mark size=3pt, mark=o, mark options={solid, mycolor2}]
  table[row sep=crcr]{%
-20	10556.9412025256\\
-5.71428571428571	5920.32901394219\\
8.57142857142857	2536.92858999593\\
22.8571428571429	1687.55470213786\\
37.1428571428571	1444.69420283069\\
51.4285714285714	1399.66381923561\\
65.7142857142857	1392.77266480979\\
80	1390.85780855451\\
};
\addlegendentry{Final RMSE under propagated orbit}

\addplot [color=mycolor3, dashed, line width=1.2pt, mark size=3pt, mark=triangle, mark options={solid, mycolor3}]
  table[row sep=crcr]{%
-20	10443.6886592821\\
-5.71428571428571	5746.83639784555\\
8.57142857142857	2094.91285412346\\
22.8571428571429	955.656946228516\\
37.1428571428571	407.292057229586\\
51.4285714285714	206.261241521568\\
65.7142857142857	141.881862081285\\
80	118.683165610598\\
};
\addlegendentry{Final  RMSE under calibrated orbit}

\addplot [color=black, line width=1pt, mark size=3pt, mark=asterisk, mark options={solid, black}]
  table[row sep=crcr]{%
-20	11553.5027697272\\
-5.71428571428571	5227.95467861996\\
8.57142857142857	2614.76513486644\\
22.8571428571429	1697.22168779289\\
37.1428571428571	1454.36406534622\\
51.4285714285714	1402.88261340511\\
65.7142857142857	1392.63279167865\\
80	1391.07246637209\\
};
\addlegendentry{LB under propagated orbit}

\addplot [color=black, line width=1pt, mark size=3pt, mark=x, mark options={solid, black}]
  table[row sep=crcr]{%
-20	11469.3722466479\\
-5.71428571428571	5039.60799902269\\
8.57142857142857	2214.38882225604\\
22.8571428571429	972.995915123519\\
37.1428571428571	427.531562759612\\
51.4285714285714	187.856114161782\\
65.7142857142857	82.5434203140275\\
80	36.2693362679373\\
};
\addlegendentry{LB under true orbit}

\addplot [color=black, line width=1pt]
  table[row sep=crcr]{%
-20	1390.41117421127\\
-5.71428571428571	1390.35919670494\\
8.57142857142857	1390.44189480243\\
22.8571428571429	1390.61426378225\\
37.1428571428571	1390.10299039152\\
51.4285714285714	1390.24770069912\\
65.7142857142857	1390.18432872282\\
80	1390.59941547663\\
};

\draw[->, thick]
(axis cs:-10,370) -- (axis cs:-10,1040);
\node[anchor=west, font=\footnotesize] at (axis cs:-20,150) {$\sqrt{\text{tr}\big([\text{Bias}(\xiv_0)]_{1:3,1:3}\big)}$ in~\eqref{eq_lbm}};

\end{axis}

\end{tikzpicture}%
    \caption{Theoretical bounds and estimation \ac{rmse} versus transmit power, including the \ac{rmse} of linear LS initialization, the final \acp{rmse} under the propagated orbit and true orbit, as well as the LBs under the propagated orbit and true orbit.
        }
    \label{fig_estimation_result}
\end{figure}

Finally, we evaluate the effectiveness of the positioning method proposed in Section~\ref{sub_sec_pos}, by comparing its RMSE with the derived positioning error bounds. We consider a scenario with \(M=4\) \acp{bs}, each providing four observations at a single observation epoch. Figure~\ref{fig_estimation_result} compares three solutions versus transmit power \(P_\mathsf{T}\in[-20,80]\) dBm: (i) the linear LS initializer in~\eqref{eq_init_close_sol}; (ii) the final solution of~\eqref{eq_pos_only} using the mismatched orbit; and (iii) the final solution of~\eqref{eq_pos_only} using the orbit calibrated by the \ac{map} estimator. The \ac{ls} initializer, computed under the erroneous orbit, improves with power, but the \ac{rmse} remains on the order of \(1\times10^4\) m. Although this initial solution is too coarse for practical positioning, it provides a useful starting point for the subsequent refinement. The solution with the mismatched orbit tracks the \ac{rmse} across the full power range, which verifies the tightness of the derived bound. After orbit correction, the estimator performance further improves and closely aligns with the ideal \ac{crb}, showing that the proposed method is capable of achieving near-optimal positioning accuracy when accurate orbital information is available.

\section{Conclusion} \label{sec_conclusion}

In this paper, we show that orbital mismatch is a fundamental factor that degrades LEO positioning performance. By characterizing orbital mismatch and incorporating it into the \ac{mcrb} analysis, we reveal that the resulting positioning error can become bias-dominated and cannot be fully mitigated by increasing transmit power or collecting more observations. To mitigate this issue, we propose that leveraging historical error statistics as prior knowledge for orbit calibration is an effective means to reduce the adverse impact of orbital errors. By incorporating these statistics into a Bayesian framework, the proposed method improves both orbit estimation and subsequent positioning performance. 

\appendices

\section{An Orbital Propagator in QuaDRiGa}\label{sec_app_orb_model}

This appendix presents an orbit propagator suggested by the QuaDRiGa platform~\cite{Jaeckel2021}. In this propagator, parameters~\(a\), \(e\), and \(i\) are treated as constants, while~\(\Omega\), \(\omega\), and \(\gamma\) are time-dependent: \(\Omega\) and \(\omega\) evolve due to the secular perturbation induced by the Earth's oblateness, while \(\gamma\) evolves as the satellite moves along the orbit. Therefore, the orbital element vector at time \(t\) is characterized by \(\ov_t = [a, e, i, \Omega_t, \omega_t, \gamma_t]^\TT\), where a subscript \(t\) is used to denote the corresponding orbital element at time \(t\).
     
    Given the Earth's non-sphericity factor~\(J_2\), the Earth's radius~\(R_e\), the Earth's mass~\(M_e\), and gravitational constant~\(\mu\), we first compute the perturbation coefficient~\(\kappa\) and the mean motion correction~\(\bar m\) as follows~\cite{Jaeckel2021}:
    \begin{gather} \label{eq_perturbation}
         \kappa = \frac{1.5 \cdot J_2 R_e^2}{a^2(1 - e^2)^2}, \\
         \bar m = \sqrt{\frac{\mu M_e}{a^3}} \left(1+\kappa(1-1.5 \cdot{\sin^2{i}} )\sqrt{1 - e^2}\right).
     \end{gather}
    The \ac{raan} and argument of periapsis at time \(t\) are then updated as
    \begin{gather}
       \Omega_t = \Omega_{t_0} - (t-t_0)\cdot \kappa\cdot \bar m\cdot\cos{i}, \\
    \omega_t = \omega_{t_0} + (t-t_0)\cdot \kappa \cdot \bar m(2-2.5\cdot{\sin^2{i}}).
    \end{gather}
    To recover the satellite position along the orbit, we further introduce the mean anomaly~\(\bar M_t\) as an auxiliary propagation variable. The mean anomaly and true anomaly are mutually related through the standard Kepler relations in~\eqref{eq_mean_anomaly}--\eqref{eq_anomaly}.
    Specifically, the mean anomaly is updated as
    \begin{align}\label{eq_mean_anomaly}
        \bar M_t = \bar M_{t_0} + \bar m(t-t_0),
    \end{align}
    where \(\bar M_{t_0}\) denotes the mean anomaly at the reference epoch \(t_0\), which is obtained from the true anomaly through the inverse Kepler relations.
    The corresponding eccentric anomaly \(E_t\) is then obtained from Kepler's equation
    \begin{equation}
        E_t - e \sin E_t = \bar M_t.
    \end{equation}
    Finally, the true anomaly at time~\(t\), i.e., $\gamma_t$, can be obtained by solving the following equation:
    \begin{equation}\label{eq_anomaly}
        \tan({\gamma_t}/{2}) = \sqrt{{(1+e)}/{(1-e)}}\tan({E_t}/{2}).
    \end{equation}

\section{Identifiability of Orbit Calibration} \label{sec_app_identify}

We note the following property of the satellite-UE geometry. For a single \ac{ue} observed at each epoch, the positioning observables are invariant in the \ac{eci} frame under any global rotation \(\Qm\in SO(3)\) applied simultaneously to the user position, the satellite state, and the satellite local frame. Consequently, the user position and the satellite state are not uniquely identifiable from such observations alone. This relationship can be justified as follows.

Consider an arbitrary rotation matrix \(\Qm\in SO(3)\). Apply this rotation to the entire geometry:
\begin{equation}
    \pv'=\Qm\pv,\qquad
    \pv_{\mathsf{s},t}'=\Qm\pv_{\mathsf{s},t},\qquad
    \vv_t'=\Qm\vv_t.
\end{equation}
Then the propagation distance remains unchanged:
\begin{equation}
    d_t'
    =
    \|\pv'-\pv_{\mathsf{s},t}'\|_2
    =
    \|\Qm(\pv-\pv_{\mathsf{s},t})\|_2
    =
    \|\pv-\pv_{\mathsf{s},t}\|_2
    =
    d_t.
\end{equation}
Next, consider the satellite local frame. Let its basis vectors be \(\xv_t,\yv_t,\zv_t\), collected in the rotation matrix \(\Rm_{\mathsf{s},t}=[\xv_t,\yv_t,\zv_t]\). Under the same global rotation, these basis vectors become
\begin{equation}
    \xv_t'=\Qm\xv_t,\qquad
    \yv_t'=\Qm\yv_t,\qquad
    \zv_t'=\Qm\zv_t,
\end{equation}
and therefore the transformed local frame is
\begin{equation}
    \Rm_{\mathsf{s},t}'=\Qm\Rm_{\mathsf{s},t}.
\end{equation}
For the \acp{aod}, we have
\begin{align}
    (\Rm_{\mathsf{s},t}')^{\TT}(\pv'-\pv_{\mathsf{s},t}')
    &=
    (\Qm\Rm_{\mathsf{s},t})^{\TT}\Qm(\pv-\pv_{\mathsf{s},t}) \notag\\
    &=
    \Rm_{\mathsf{s},t}^{\TT}\Qm^{\TT}\Qm(\pv-\pv_{\mathsf{s},t}) =
    \Rm_{\mathsf{s},t}^{\TT}(\pv-\pv_{\mathsf{s},t}).
\end{align}
Hence, both azimuth and elevation remain unchanged. For the propagation delay,
\begin{equation}
    \tau_t'
    =
    \frac{\|\pv'-\pv_{\mathsf{s},t}'\|_2}{c}+\Delta
    =
    \frac{\|\pv-\pv_{\mathsf{s},t}\|_2}{c}+\Delta
    =
    \tau_t.
\end{equation}
For the normalized Doppler shift,
\begin{align}
    \nu_t'
    &=
    \frac{(\vv_t')^{\TT}(\pv'-\pv_{\mathsf{s},t}')}
    {c\|\pv'-\pv_{\mathsf{s},t}'\|_2} =
    \frac{(\Qm\vv_t)^{\TT}\Qm(\pv-\pv_{\mathsf{s},t})}
    {c\|\pv-\pv_{\mathsf{s},t}\|_2} \notag\\
    &=
    \frac{\vv_t^{\TT}\Qm^{\TT}\Qm(\pv-\pv_{\mathsf{s},t})}
    {c\|\pv-\pv_{\mathsf{s},t}\|_2} =
    \frac{\vv_t^{\TT}(\pv-\pv_{\mathsf{s},t})}
    {c\|\pv-\pv_{\mathsf{s},t}\|_2}
    =
    \nu_t.
\end{align}
Therefore, when the same global rotation \(\Qm\) is applied to all epochs, the entire multi-epoch observation set remains unchanged as well. This proves that infinitely many rotated user-satellite states generate the same observation set. Thus, the stand-alone calibration problem is not uniquely identifiable from a single \ac{ue} alone.

The above derivation shows that the stand-alone orbit-calibration problem with a single \ac{ue} is fundamentally non-identifiable. Without external references, the user state and the satellite state cannot be uniquely determined from the available observations. To remove this ambiguity, additional known references must be introduced.

\bibliographystyle{IEEEtran}
\bibliography{references}
\end{document}